\documentstyle[12pt,aaspp4]{article}
\def\PsfigVersion{1.10}
\def\setDriver{\DvipsDriver} 
\ifx\undefined\psfig\else \fi
\let\LaTeXAtSign=\@
\let\@=\relax
\edef\psfigRestoreAt{\catcode`\@=\number\catcode`@\relax}
\catcode`\@=11\relax
\newwrite\@unused
\def\ps@typeout#1{{\let\protect\string\immediate\write\@unused{#1}}}

\def\DvipsDriver{
	\ps@typeout{psfig/tex \PsfigVersion -dvips}
\def\PsfigSpecials{\DvipsSpecials} 	\def\ps@dir{/}
\def\ps@predir{} }
\def\OzTeXDriver{
	\ps@typeout{psfig/tex \PsfigVersion -oztex}
	\def\PsfigSpecials{\OzTeXSpecials}
	\def\ps@dir{:}
	\def\ps@predir{:}
	\catcode`\^^J=5
}

\def\figurepath{./:}

\def\DoPaths#1{\expandafter\EachPath#1\stoplist}
\def\leer{}
\def\EachPath#1:#2\stoplist{
  \ExistsFile{#1}{\SearchedFile}
  \ifx#2\leer
  \else
    \expandafter\EachPath#2\stoplist
  \fi}
\def\ps@dir{/}
\def\ExistsFile#1#2{%
   \openin1=\ps@predir#1\ps@dir#2
   \ifeof1
       \closein1
   \else
       \closein1
        \ifx\ps@founddir\leer
           \edef\ps@founddir{#1}
        \fi
   \fi}
\def\get@dir#1{%
  \def\ps@founddir{}
  \def\SearchedFile{#1}
  \DoPaths\figurepath
}

\def\@nnil{\@nil}
\def\@empty{}
\def\@psdonoop#1\@@#2#3{}
\def\@psdo#1:=#2\do#3{\edef\@psdotmp{#2}\ifx\@psdotmp\@empty \else
    \expandafter\@psdoloop#2,\@nil,\@nil\@@#1{#3}\fi}
\def\@psdoloop#1,#2,#3\@@#4#5{\def#4{#1}\ifx #4\@nnil \else
       #5\def#4{#2}\ifx #4\@nnil \else#5\@ipsdoloop #3\@@#4{#5}\fi\fi}
\def\@ipsdoloop#1,#2\@@#3#4{\def#3{#1}\ifx #3\@nnil 
       \let\@nextwhile=\@psdonoop \else
      #4\relax\let\@nextwhile=\@ipsdoloop\fi\@nextwhile#2\@@#3{#4}}
\def\@tpsdo#1:=#2\do#3{\xdef\@psdotmp{#2}\ifx\@psdotmp\@empty \else
    \@tpsdoloop#2\@nil\@nil\@@#1{#3}\fi}
\def\@tpsdoloop#1#2\@@#3#4{\def#3{#1}\ifx #3\@nnil 
       \let\@nextwhile=\@psdonoop \else
      #4\relax\let\@nextwhile=\@tpsdoloop\fi\@nextwhile#2\@@#3{#4}}
\ifx\undefined\fbox
\newdimen\fboxrule
\newdimen\fboxsep
\newdimen\ps@tempdima
\newbox\ps@tempboxa
\long\def\fbox#1{\leavevmode\setbox\ps@tempboxa\hbox{#1}\ps@tempdima\fboxrule
    \advance\ps@tempdima \fboxsep \advance\ps@tempdima \dp\ps@tempboxa
   \hbox{\lower \ps@tempdima\hbox
  {\vbox{\hrule height \fboxrule
          \hbox{\vrule width \fboxrule \hskip\fboxsep
          \vbox{\vskip\fboxsep \box\ps@tempboxa\vskip\fboxsep}\hskip 
                 \fboxsep\vrule width \fboxrule}
                 \hrule height \fboxrule}}}}
\fi
\newread\ps@stream
\newif\ifnot@eof       
\newif\if@noisy        
\newif\if@atend        
\newif\if@psfile       
{\catcode`\%=12\global\gdef\epsf@start{
\def\epsf@PS{PS}
\def\epsf@getbb#1{%
\openin\ps@stream=\ps@predir#1
\ifeof\ps@stream\ps@typeout{Error, File #1 not found}\else
   {\not@eoftrue \chardef\other=12
    \def\do##1{\catcode`##1=\other}\dospecials \catcode`\ =10
    \loop
       \if@psfile
	  \read\ps@stream to \epsf@fileline
       \else{
	  \obeyspaces
          \read\ps@stream to \epsf@tmp\global\let\epsf@fileline\epsf@tmp}
       \fi
       \ifeof\ps@stream\not@eoffalse\else
       \if@psfile\else
       \expandafter\epsf@test\epsf@fileline:. \\%
       \fi
          \expandafter\epsf@aux\epsf@fileline:. \\%
       \fi
   \ifnot@eof\repeat
   }\closein\ps@stream\fi}%
\long\def\epsf@test#1#2#3:#4\\{\def\epsf@testit{#1#2}
			\ifx\epsf@testit\epsf@start\else
\ps@typeout{Warning! File does not start with `\epsf@start'.  It may not be a PostScript file.}
			\fi
			\@psfiletrue} 
{\catcode`\%=12\global\let\epsf@percent=
\long\def\epsf@aux#1#2:#3\\{\ifx#1\epsf@percent
   \def\epsf@testit{#2}\ifx\epsf@testit\epsf@bblit
	\@atendfalse
        \epsf@atend #3 . \\%
	\if@atend	
	   \if@verbose{
		\ps@typeout{psfig: found `(atend)'; continuing search}
	   }\fi
        \else
        \epsf@grab #3 . . . \\%
        \not@eoffalse
        \global\no@bbfalse
        \fi
   \fi\fi}%
\def\epsf@grab #1 #2 #3 #4 #5\\{%
   \global\def\epsf@llx{#1}\ifx\epsf@llx\empty
      \epsf@grab #2 #3 #4 #5 .\\\else
   \global\def\epsf@lly{#2}%
   \global\def\epsf@urx{#3}\global\def\epsf@ury{#4}\fi}%
\def\epsf@atendlit{(atend)} 
\def\epsf@atend #1 #2 #3\\{%
   \def\epsf@tmp{#1}\ifx\epsf@tmp\empty
      \epsf@atend #2 #3 .\\\else
   \ifx\epsf@tmp\epsf@atendlit\@atendtrue\fi\fi}

\chardef\psletter = 11 
\chardef\other = 12

\newif \ifdebug 
\newif\ifc@mpute 
\c@mputetrue 

\let\then = \relax
\def\r@dian{pt }
\let\r@dians = \r@dian
\let\dimensionless@nit = \r@dian
\let\dimensionless@nits = \dimensionless@nit
\def\internal@nit{sp }
\let\internal@nits = \internal@nit
\newif\ifstillc@nverging
\def \Mess@ge #1{\ifdebug \then \message {#1} \fi}

{ 
	\catcode `\@ = \psletter
	\gdef \nodimen {\expandafter \n@dimen \the \dimen}
	\gdef \term #1 #2 #3%
	       {\edef \t@ {\the #1}
		\edef \t@@ {\expandafter \n@dimen \the #2\r@dian}%
		\t@rm {\t@} {\t@@} {#3}%
	       }
	\gdef \t@rm #1 #2 #3%
	       {{%
		\count 0 = 0
		\dimen 0 = 1 \dimensionless@nit
		\dimen 2 = #2\relax
		\Mess@ge {Calculating term #1 of \nodimen 2}%
		\loop
		\ifnum	\count 0 < #1
		\then	\advance \count 0 by 1
			\Mess@ge {Iteration \the \count 0 \space}%
			\Multiply \dimen 0 by {\dimen 2}%
			\Mess@ge {After multiplication, term = \nodimen 0}%
			\Divide \dimen 0 by {\count 0}%
			\Mess@ge {After division, term = \nodimen 0}%
		\repeat
		\Mess@ge {Final value for term #1 of 
				\nodimen 2 \space is \nodimen 0}%
		\xdef \Term {#3 = \nodimen 0 \r@dians}%
		\aftergroup \Term
	       }}
	\catcode `\p = \other
	\catcode `\t = \other
	\gdef \n@dimen #1pt{#1} 
}

\def \Divide #1by #2{\divide #1 by #2} 

\def \Multiply #1by #2
       {{
	\count 0 = #1\relax
	\count 2 = #2\relax
	\count 4 = 65536
	\Mess@ge {Before scaling, count 0 = \the \count 0 \space and
			count 2 = \the \count 2}%
	\ifnum	\count 0 > 32767 
	\then	\divide \count 0 by 4
		\divide \count 4 by 4
	\else	\ifnum	\count 0 < -32767
		\then	\divide \count 0 by 4
			\divide \count 4 by 4
		\else
		\fi
	\fi
	\ifnum	\count 2 > 32767 
	\then	\divide \count 2 by 4
		\divide \count 4 by 4
	\else	\ifnum	\count 2 < -32767
		\then	\divide \count 2 by 4
			\divide \count 4 by 4
		\else
		\fi
	\fi
	\multiply \count 0 by \count 2
	\divide \count 0 by \count 4
	\xdef \product {#1 = \the \count 0 \internal@nits}%
	\aftergroup \product
       }}

\def\r@duce{\ifdim\dimen0 > 90\r@dian \then   
		\multiply\dimen0 by -1
		\advance\dimen0 by 180\r@dian
		\r@duce
	    \else \ifdim\dimen0 < -90\r@dian \then  
		\advance\dimen0 by 360\r@dian
		\r@duce
		\fi
	    \fi}

\def\Sine#1%
       {{%
	\dimen 0 = #1 \r@dian
	\r@duce
	\ifdim\dimen0 = -90\r@dian \then
	   \dimen4 = -1\r@dian
	   \c@mputefalse
	\fi
	\ifdim\dimen0 = 90\r@dian \then
	   \dimen4 = 1\r@dian
	   \c@mputefalse
	\fi
	\ifdim\dimen0 = 0\r@dian \then
	   \dimen4 = 0\r@dian
	   \c@mputefalse
	\fi
	\ifc@mpute \then
		\divide\dimen0 by 180
		\dimen0=3.141592654\dimen0
		\dimen 2 = 3.1415926535897963\r@dian 
		\divide\dimen 2 by 2 
		\Mess@ge {Sin: calculating Sin of \nodimen 0}%
		\count 0 = 1 
		\dimen 2 = 1 \r@dian 
		\dimen 4 = 0 \r@dian 
		\loop
			\ifnum	\dimen 2 = 0 
			\then	\stillc@nvergingfalse 
			\else	\stillc@nvergingtrue
			\fi
			\ifstillc@nverging 
			\then	\term {\count 0} {\dimen 0} {\dimen 2}%
				\advance \count 0 by 2
				\count 2 = \count 0
				\divide \count 2 by 2
				\ifodd	\count 2 
				\then	\advance \dimen 4 by \dimen 2
				\else	\advance \dimen 4 by -\dimen 2
				\fi
		\repeat
	\fi		
			\xdef \sine {\nodimen 4}%
       }}

\def\Cosine#1{\ifx\sine\UnDefined\edef\Savesine{\relax}\else
		             \edef\Savesine{\sine}\fi
	{\dimen0=#1\r@dian\advance\dimen0 by 90\r@dian
	 \Sine{\nodimen 0}
	 \xdef\cosine{\sine}
	 \xdef\sine{\Savesine}}}	      

\def\psdraft{
	\def\@psdraft{0}
}
\def\psfull{
	\def\@psdraft{100}
}

\psfull

\newif\if@scalefirst
\def\psscalefirst{\@scalefirsttrue}
\def\psrotatefirst{\@scalefirstfalse}
\psrotatefirst

\newif\if@draftbox
\def\psnodraftbox{
	\@draftboxfalse
}
\def\psdraftbox{
	\@draftboxtrue
}
\@draftboxtrue

\newif\if@prologfile
\newif\if@postlogfile
\def\pssilent{
	\@noisyfalse
}
\def\psnoisy{
	\@noisytrue
}
\psnoisy
\newif\if@bbllx
\newif\if@bblly
\newif\if@bburx
\newif\if@bbury
\newif\if@height
\newif\if@width
\newif\if@rheight
\newif\if@rwidth
\newif\if@angle
\newif\if@clip
\newif\if@verbose
\def\@p@@sclip#1{\@cliptrue}
\newif\if@decmpr
\def\@p@@sfigure#1{\def\@p@sfile{null}\def\@p@sbbfile{null}\@decmprfalse
   \openin1=\ps@predir#1
   \ifeof1
	\closein1
	\get@dir{#1}
	\ifx\ps@founddir\leer
		\openin1=\ps@predir#1.bb
		\ifeof1
			\closein1
			\get@dir{#1.bb}
			\ifx\ps@founddir\leer
				\ps@typeout{Can't find #1 in \figurepath}
			\else
				\@decmprtrue
				\def\@p@sfile{\ps@founddir\ps@dir#1}
				\def\@p@sbbfile{\ps@founddir\ps@dir#1.bb}
			\fi
		\else
			\closein1
			\@decmprtrue
			\def\@p@sfile{#1}
			\def\@p@sbbfile{#1.bb}
		\fi
	\else
		\def\@p@sfile{\ps@founddir\ps@dir#1}
		\def\@p@sbbfile{\ps@founddir\ps@dir#1}
	\fi
   \else
	\closein1
	\def\@p@sfile{#1}
	\def\@p@sbbfile{#1}
   \fi
}
\def\@p@@sfile#1{\@p@@sfigure{#1}}
\def\@p@@sbbllx#1{
		\@bbllxtrue
		\dimen100=#1
		\edef\@p@sbbllx{\number\dimen100}
}
\def\@p@@sbblly#1{
		\@bbllytrue
		\dimen100=#1
		\edef\@p@sbblly{\number\dimen100}
}
\def\@p@@sbburx#1{
		\@bburxtrue
		\dimen100=#1
		\edef\@p@sbburx{\number\dimen100}
}
\def\@p@@sbbury#1{
		\@bburytrue
		\dimen100=#1
		\edef\@p@sbbury{\number\dimen100}
}
\def\@p@@sheight#1{
		\@heighttrue
		\dimen100=#1
   		\edef\@p@sheight{\number\dimen100}
}
\def\@p@@swidth#1{
		\@widthtrue
		\dimen100=#1
		\edef\@p@swidth{\number\dimen100}
}
\def\@p@@srheight#1{
		\@rheighttrue
		\dimen100=#1
		\edef\@p@srheight{\number\dimen100}
}
\def\@p@@srwidth#1{
		\@rwidthtrue
		\dimen100=#1
		\edef\@p@srwidth{\number\dimen100}
}
\def\@p@@sangle#1{
		\@angletrue
		\edef\@p@sangle{#1} 
}
\def\@p@@ssilent#1{ 
		\@verbosefalse
}
\def\@p@@sprolog#1{\@prologfiletrue\def\@prologfileval{#1}}
\def\@p@@spostlog#1{\@postlogfiletrue\def\@postlogfileval{#1}}
\def\@cs@name#1{\csname #1\endcsname}
\def\@setparms#1=#2,{\@cs@name{@p@@s#1}{#2}}
\def\ps@init@parms{
		\@bbllxfalse \@bbllyfalse
		\@bburxfalse \@bburyfalse
		\@heightfalse \@widthfalse
		\@rheightfalse \@rwidthfalse
		\def\@p@sbbllx{}\def\@p@sbblly{}
		\def\@p@sbburx{}\def\@p@sbbury{}
		\def\@p@sheight{}\def\@p@swidth{}
		\def\@p@srheight{}\def\@p@srwidth{}
		\def\@p@sangle{0}
		\def\@p@sfile{} \def\@p@sbbfile{}
		\def\@p@scost{10}
		\def\@sc{}
		\@prologfilefalse
		\@postlogfilefalse
		\@clipfalse
		\if@noisy
			\@verbosetrue
		\else
			\@verbosefalse
		\fi
}
\def\parse@ps@parms#1{
	 	\@psdo\@psfiga:=#1\do
		   {\expandafter\@setparms\@psfiga,}}
\newif\ifno@bb
\def\bb@missing{
	\if@verbose{
		\ps@typeout{psfig: searching \@p@sbbfile \space  for bounding box}
	}\fi
	\no@bbtrue
	\epsf@getbb{\@p@sbbfile}
        \ifno@bb \else \bb@cull\epsf@llx\epsf@lly\epsf@urx\epsf@ury\fi
}	
\def\bb@cull#1#2#3#4{
	\dimen100=#1 bp\edef\@p@sbbllx{\number\dimen100}
	\dimen100=#2 bp\edef\@p@sbblly{\number\dimen100}
	\dimen100=#3 bp\edef\@p@sbburx{\number\dimen100}
	\dimen100=#4 bp\edef\@p@sbbury{\number\dimen100}
	\no@bbfalse
}
\newdimen\p@intvaluex
\newdimen\p@intvaluey
\def\rotate@#1#2{{\dimen0=#1 sp\dimen1=#2 sp
		  \global\p@intvaluex=\cosine\dimen0
		  \dimen3=\sine\dimen1
		  \global\advance\p@intvaluex by -\dimen3
		  \global\p@intvaluey=\sine\dimen0
		  \dimen3=\cosine\dimen1
		  \global\advance\p@intvaluey by \dimen3
		  }}
\def\compute@bb{
		\no@bbfalse
		\if@bbllx \else \no@bbtrue \fi
		\if@bblly \else \no@bbtrue \fi
		\if@bburx \else \no@bbtrue \fi
		\if@bbury \else \no@bbtrue \fi
		\ifno@bb \bb@missing \fi
		\ifno@bb \ps@typeout{FATAL ERROR: no bb supplied or found}
			\no-bb-error
		\fi
		\count203=\@p@sbburx
		\count204=\@p@sbbury
		\advance\count203 by -\@p@sbbllx
		\advance\count204 by -\@p@sbblly
		\edef\ps@bbw{\number\count203}
		\edef\ps@bbh{\number\count204}
		\if@angle 
			\Sine{\@p@sangle}\Cosine{\@p@sangle}
	        	{\dimen100=\maxdimen\xdef\r@p@sbbllx{\number\dimen100}
					    \xdef\r@p@sbblly{\number\dimen100}
			                    \xdef\r@p@sbburx{-\number\dimen100}
					    \xdef\r@p@sbbury{-\number\dimen100}}
                        \def\minmaxtest{
			   \ifnum\number\p@intvaluex<\r@p@sbbllx
			      \xdef\r@p@sbbllx{\number\p@intvaluex}\fi
			   \ifnum\number\p@intvaluex>\r@p@sbburx
			      \xdef\r@p@sbburx{\number\p@intvaluex}\fi
			   \ifnum\number\p@intvaluey<\r@p@sbblly
			      \xdef\r@p@sbblly{\number\p@intvaluey}\fi
			   \ifnum\number\p@intvaluey>\r@p@sbbury
			      \xdef\r@p@sbbury{\number\p@intvaluey}\fi
			   }
			\rotate@{\@p@sbbllx}{\@p@sbblly}
			\minmaxtest
			\rotate@{\@p@sbbllx}{\@p@sbbury}
			\minmaxtest
			\rotate@{\@p@sbburx}{\@p@sbblly}
			\minmaxtest
			\rotate@{\@p@sbburx}{\@p@sbbury}
			\minmaxtest
			\edef\@p@sbbllx{\r@p@sbbllx}\edef\@p@sbblly{\r@p@sbblly}
			\edef\@p@sbburx{\r@p@sbburx}\edef\@p@sbbury{\r@p@sbbury}
		\fi
		\count203=\@p@sbburx
		\count204=\@p@sbbury
		\advance\count203 by -\@p@sbbllx
		\advance\count204 by -\@p@sbblly
		\edef\@bbw{\number\count203}
		\edef\@bbh{\number\count204}
}
\def\in@hundreds#1#2#3{\count240=#2 \count241=#3
		     \count100=\count240	
		     \divide\count100 by \count241
		     \count101=\count100
		     \multiply\count101 by \count241
		     \advance\count240 by -\count101
		     \multiply\count240 by 10
		     \count101=\count240	
		     \divide\count101 by \count241
		     \count102=\count101
		     \multiply\count102 by \count241
		     \advance\count240 by -\count102
		     \multiply\count240 by 10
		     \count102=\count240	
		     \divide\count102 by \count241
		     \count200=#1\count205=0
		     \count201=\count200
			\multiply\count201 by \count100
		 	\advance\count205 by \count201
		     \count201=\count200
			\divide\count201 by 10
			\multiply\count201 by \count101
			\advance\count205 by \count201
		     \count201=\count200
			\divide\count201 by 100
			\multiply\count201 by \count102
			\advance\count205 by \count201
		     \edef\@result{\number\count205}
}
\def\compute@wfromh{
		\in@hundreds{\@p@sheight}{\@bbw}{\@bbh}
		\edef\@p@swidth{\@result}
}
\def\compute@hfromw{
	        \in@hundreds{\@p@swidth}{\@bbh}{\@bbw}
		\edef\@p@sheight{\@result}
}
\def\compute@handw{
		\if@height 
			\if@width
			\else
				\compute@wfromh
			\fi
		\else 
			\if@width
				\compute@hfromw
			\else
				\edef\@p@sheight{\@bbh}
				\edef\@p@swidth{\@bbw}
			\fi
		\fi
}
\def\compute@resv{
		\if@rheight \else \edef\@p@srheight{\@p@sheight} \fi
		\if@rwidth \else \edef\@p@srwidth{\@p@swidth} \fi
}
\def\compute@sizes{
	\compute@bb
	\if@scalefirst\if@angle
	\if@width
	   \in@hundreds{\@p@swidth}{\@bbw}{\ps@bbw}
	   \edef\@p@swidth{\@result}
	\fi
	\if@height
	   \in@hundreds{\@p@sheight}{\@bbh}{\ps@bbh}
	   \edef\@p@sheight{\@result}
	\fi
	\fi\fi
	\compute@handw
	\compute@resv}
\def\OzTeXSpecials{
	\special{empty.ps /@isp {true} def}
	\special{empty.ps \@p@swidth \space \@p@sheight \space
			\@p@sbbllx \space \@p@sbblly \space
			\@p@sbburx \space \@p@sbbury \space
			startTexFig \space }
	\if@clip{
		\if@verbose{
			\ps@typeout{(clip)}
		}\fi
		\special{empty.ps doclip \space }
	}\fi
	\if@angle{
		\if@verbose{
			\ps@typeout{(rotate)}
		}\fi
		\special {empty.ps \@p@sangle \space rotate \space} 
	}\fi
	\if@prologfile
	    \special{\@prologfileval \space } \fi
	\if@decmpr{
		\if@verbose{
			\ps@typeout{psfig: Compression not available
			in OzTeX version \space }
		}\fi
	}\else{
		\if@verbose{
			\ps@typeout{psfig: including \@p@sfile \space }
		}\fi
		\special{epsf=\ps@predir\@p@sfile \space }
	}\fi
	\if@postlogfile
	    \special{\@postlogfileval \space } \fi
	\special{empty.ps /@isp {false} def}
}
\def\DvipsSpecials{
	\special{ps::[begin] 	\@p@swidth \space \@p@sheight \space
			\@p@sbbllx \space \@p@sbblly \space
			\@p@sbburx \space \@p@sbbury \space
			startTexFig \space }
	\if@clip{
		\if@verbose{
			\ps@typeout{(clip)}
		}\fi
		\special{ps:: doclip \space }
	}\fi
	\if@angle
		\if@verbose{
			\ps@typeout{(clip)}
		}\fi
		\special {ps:: \@p@sangle \space rotate \space} 
	\fi
	\if@prologfile
	    \special{ps: plotfile \@prologfileval \space } \fi
	\if@decmpr{
		\if@verbose{
			\ps@typeout{psfig: including \@p@sfile.Z \space }
		}\fi
		\special{ps: plotfile "`zcat \@p@sfile.Z" \space }
	}\else{
		\if@verbose{
			\ps@typeout{psfig: including \@p@sfile \space }
		}\fi
		\special{ps: plotfile \@p@sfile \space }
	}\fi
	\if@postlogfile
	    \special{ps: plotfile \@postlogfileval \space } \fi
	\special{ps::[end] endTexFig \space }
}
\def\psfig#1{\vbox {
	\ps@init@parms
	\parse@ps@parms{#1}
	\compute@sizes
	\ifnum\@p@scost<\@psdraft{
		\PsfigSpecials 
		\vbox to \@p@srheight sp{
			\hbox to \@p@srwidth sp{
				\hss
			}
		\vss
		}
	}\else{
		\if@draftbox{		
			\hbox{\fbox{\vbox to \@p@srheight sp{
			\vss
			\hbox to \@p@srwidth sp{ \hss 
			 \hss }
			\vss
			}}}
		}\else{
			\vbox to \@p@srheight sp{
			\vss
			\hbox to \@p@srwidth sp{\hss}
			\vss
			}
		}\fi

	}\fi
}}
\psfigRestoreAt
\setDriver
\let\@=\LaTeXAtSign

\def\deg{\ifmmode ^{\circ}
         \else $^{\circ}$\fi}
\def\pdeg{\ifmmode $\setbox0=\hbox{$^{\circ}$}\rlap{\hskip.11\wd0 .}$^{\circ}
          \else \setbox0=\hbox{$^{\circ}$}\rlap{\hskip.11\wd0 .}$^{\circ}$\fi}
\def\arcs{\ifmmode {^{\scriptscriptstyle\prime\prime}}
          \else $^{\scriptscriptstyle\prime\prime}$\fi}
\def\arcm{\ifmmode {^{\scriptscriptstyle\prime}}
          \else $^{\scriptscriptstyle\prime}$\fi}
\newdimen\sa  \newdimen\sb
\def\parcs{\sa=.07em \sb=.03em
     \ifmmode $\rlap{.}$^{\scriptscriptstyle\prime\kern -\sb\prime}$\kern -\sa$
     \else \rlap{.}$^{\scriptscriptstyle\prime\kern -\sb\prime}$\kern -\sa\fi}
\def\parcm{\sa=.08em \sb=.03em
     \ifmmode $\rlap{.}\kern\sa$^{\scriptscriptstyle\prime}$\kern-\sb$
     \else \rlap{.}\kern\sa$^{\scriptscriptstyle\prime}$\kern-\sb\fi}
\def\gtorder{\mathrel{\raise.3ex\hbox{$>$}\mkern-14mu
             \lower0.6ex\hbox{$\sim$}}}
\def\ltorder{\mathrel{\raise.3ex\hbox{$<$}\mkern-14mu
             \lower0.6ex\hbox{$\sim$}}}
\newcommand{\kms}{\mbox{ km~s$^{-1}$}}

\begin{document}

\title{The Flat Spectrum Radio Luminosity Function, \protect\\
       Gravitational Lensing, Galaxy Ellipticities and \protect\\
       Cosmology}
\author{C.S. Kochanek}
\affil{Harvard-Smithsonian Center for Astrophysics, MS-51\protect \\ 
       60 Garden Street \protect \\
       Cambridge MA 02138 }
\authoremail{ckochanek@cfa.harvard.edu}

\begin{abstract}
The number of lenses found in the JVAS survey of flat-spectrum radio sources for
gravitational lenses is consistent with statistical models of optical surveys for
lensed quasars. The 90\% confidence limit on $\Omega_0$ in flat cosmological models 
($\Omega_0+\lambda_0=1$) is approximately $0.15 \ltorder \Omega_0 \ltorder 2$.  
Depending on the RLF model, we predict 2.4 to 3.6 lenses in the JVAS survey 
and in the first part of the fainter CLASS survey, and 0.3 to 0.6 
lenses in the brighter PHFS survey for an $\Omega_0=1$ model.  The uncertainties 
are due to the small numbers of lenses (there are only 4 compact JVAS lenses) 
and the uncertainties in the radio luminosity function (RLF) caused by the lack of 
information on the redshift distribution of 10-300 mJy radio sources.  
If we force the models to produce the observed number of JVAS lenses,
the mean redshift of a 50 mJy source varies
from $z_s=0.4$ for $\Omega_0=0$, to $1.9$ for $\Omega_0=1$, to almost $4.0$ 
for $\Omega_0=2$ when $\Omega_0+\lambda_0=1$.
The source fluxes and redshifts of the lenses in the JVAS and CLASS surveys are 
consistent with the statistical models.  
The numbers of four-image lenses found in the JVAS survey and in surveys for
lensed quasars are mutually consistent, but slightly larger than expected for models 
using the observed axis ratios of E and S0 galaxies.  
The best fits to the lens data require a projected axis ratio of $b/a=0.50$ with
a 90\% confidence range of $0.25 < b/a < 0.65$.  
\end{abstract}


\def\dv{\Delta V}
\def\df{\Delta {\cal F}}

\section{Introduction}

The number of gravitational lenses found in systematic surveys for lenses is a
strong constraint on cosmological models, particularly models with a large 
cosmological constant (Turner 1990, Fukugita, Futamase \& Kasai 1990).  
Quantitative analyses of surveys for
multiply imaged quasars (Kochanek 1993, 1996, Maoz \& Rix 1993)
currently give a formal two standard deviation (2-$\sigma$) upper
limit on the cosmological constant of $\lambda_0 < 0.66$, effectively ruling
out values large enough to be cosmologically interesting.  The published
optical samples contain 862 quasars and 5 lenses produced by galaxies.  
The $\lambda_0$ limit includes all the statistical uncertainties in the model 
(the number of lenses,
galaxy luminosity function, dynamical normalizations and the quasar number
counts), and is insensitive to the mass distribution of the lens galaxies.
Galaxy evolution weakly affects the limit (Mao 1991, Mao \& Kochanek 1994,
Rix et al. 1994) because physical models of galaxy mergers generally preserve
the expected number of lenses. Moreover, gravitational lensing depends 
on the properties of low redshift elliptical galaxies ($z < 1$) even if
$\lambda_0=1$, and recent observations show little evolution in this population 
(e.g. Lilly et al. 1995, Dickinson 1995, Driver et al. 
1995).  For bright quasars the selection effects of the survey to
find the quasars are probably unimportant (Kochanek 1991, 1996), although 
this issue should be reexamined.   Kochanek (1993) 
found no detectable bias in the lens sample between radio, color,
and spectrally selected quasars. Absorption by
dust in the lens galaxies can strongly affect the statistical models
(Kochanek 1991, 1996, Tomita 1995), but the current dust content
of E/S0 galaxies is far too low to alter the statistics. 
Nonetheless, absorption can be a large systematic problem
in the optically selected lens samples, and there is evidence in some
of the radio lenses (e.g. Lawrence et al. 1994) for high extinction.

Most of the known lenses produced by galaxies were not found by surveying 
quasars at optical wavelengths, but by surveying radio sources.  The 
largest number have been found in the MIT/Greenbank or MG survey (Burke et al.
1992), which has found six lenses. The MG survey examines all radio 
sources exceeding a flux limit of 50 mJy (see Leh\'ar 1991), and the 
resulting lens sample is dominated by extended steep spectrum sources 
that form radio ring lenses such as MG1131+045 (Hewitt et al.  1988).  
The extended structure of steep spectrum sources complicates statistical 
analyses because the finite size of the sources strongly modifies the 
lensing probability (Kochanek \& Lawrence 1990).  
The Jodrell Bank-VLA Astrometric Survey (JVAS, Patnaik 1994, Patnaik et al. 
1992, King et al. 1996, King \& Browne 1996), CLASS (Jackson et al. 1995,
Myers et al. 1995, Myers 1996), and PHFS (Webster et al. 1996) surveys 
are restricted
to flat spectrum radio sources.  Flat spectrum sources are generally compact
radio cores, making it much simpler to recognize lensed systems and to 
compute their statistics.  The JVAS sample consists
of 2200 sources brighter than 200 mJy.  It contains five lenses, two doubles
(B 0218+357 and a candidate), two quads (MG 0414+0534 and B 1422+231), and
one hybrid (B 1938+666).  The published subset of the CLASS survey has 676 
sources brighter than 25 mJy and 2575 sources brighter than 50 mJy.  It 
contains at least one double (CLASS 1600+434)
and one quad (CLASS 1608+656).  The PHFS survey has 323 sources brighter than
500 mJy, and no lensing results are known.

There are many advantages to studying the statistics of radio lens surveys
rather than optical quasar lens surveys.  The radio samples are selected purely 
on their radio flux and spectral index, so there is almost no imaginable bias 
against including lenses in the sample.  The radio samples are immune to the 
effects of dust extinction in lens galaxies.  The radio surveys are
very uniform and typically have better resolution than quasar surveys.
In short, the radio surveys avoid many of the systematic errors that
may bias the conclusions of the statistical models for lensed quasars.
The redshift and flux distributions of the radio samples are 
different from that of the quasar samples, so statistical consistency 
between the radio and quasar lens samples is a good check of the 
reliability of statistical models.    
The only systematic errors common to both the radio and quasar surveys
are the number and mass distributions of lens galaxies. Both of these systematic
errors are best addressed by finding larger numbers of lenses.

The radio sources are not a panacea of course.  The problem for the radio 
surveys is the relative paucity (compared to quasar samples) of information 
on the luminosity function of radio sources as a function of redshift.  
While the number counts of radio sources are well determined (see reviews
by Dunlop 1994, Windhorst, Mathis \& Neuschaeffer 1990, Kellerman \& Wall 
1987), complete redshift
surveys exist only for bright radio sources ($S_5 \gtorder 1$ Jy), although there
are moderately complete surveys for somewhat fainter sources ($S_5 \gtorder 0.1$ Jy).
Since most of the radio lens surveys use flux limits of 25-500 mJy, the
source of a typical lens comes from the flux ranges where there is
little redshift information.  Thus the dominant uncertainty in statistical models
of radio lenses is the redshift distribution or luminosity function of
the sources.  An important goal of this study is to determine the observations 
that can eliminate this source of uncertainty in the lens calculations.  

King et al. (1996) and King \& Browne (1996)
pointed out that the numbers of four-image lens systems in the
JVAS sample is anomalously high.  Where their models, based on the Dunlop \& Peacock
(1990) RLF, predicted that 1 in 6 lenses would be a four-image lens, the JVAS sample
actually contains 2 two-image lenses, 2 four-image lenses, and one hybrid.  An 
examination of elliptical models for quasar surveys (Kochanek 1991, Wallington \& Narayan 1993,
Kasiola \& Kovner 1993) also suggests the need for very elliptical models
to make four-image systems dominate the sample of bright quasar lenses, and
models of individual lenses (e.g. Kochanek 1991, Ratnatunga et al. 1995) can also be very
elliptical.  These are all aspects of what Schechter (1996) refers to as the 
``ellipticity crisis'' in gravitational lensing.  The only plausible solutions
suggested to the problem have been that it is a statistical fluke, that is caused 
by systematic errors in lens and selection effects models, or that the extra shear is induced by
potential fluctuations along the path from the observer to the source.  Calculations 
by Bar-Kana (1996) show that the extra shear generated by large scale 
structure may be large enough to produce the observed effects, although more detailed 
calculations are needed. 

Most of this study is devoted to methods for determining the radio luminosity
function (RLF) of flat spectrum radio sources so that we can probe the effects
of its uncertainties on the statistics of gravitational lenses.  
In \S2 we summarize the data on radio number counts and redshift surveys, and
describe the statistical methods used to fit the RLF to the data.  
In \S3 we discuss the lensing calculation and develop 
models for the statistics of elliptical (density) singular isothermal
spheres.  In \S4 we explore the RLF, lens statistics, and cosmology for flat
models with a cosmological constant, considering only the total numbers of
lenses and not their relative morphologies.  In \S5 we explore the
effects of forcing the models to produce exactly the number of lenses
observed in the JVAS survey.  In \S6 we consider the 
relative numbers of two- and four-image lenses expected in the surveys and
quantify the extent of the ``ellipticity crisis''.
In \S7 we summarize our conclusions
and discuss the data required to reduce the uncertainties.

\def\hjy{\hbox{ Jy}}
\section{The Radio Luminosity Function}

In this section we summarize the constraints on the RLF of flat spectrum sources
and develop a computational method for determining the RLF using linear
regularization.  If $S_\nu \propto \nu^{-\alpha}$, then flat spectrum is 
defined by $\alpha < 0.5$, and the spectral index $\alpha$ for the sources 
in the lens surveys was usually measured between 2.7~GHz and 5~GHz.
The RLF is a two-dimensional function of redshift and
flux or luminosity, and fully determining the RLF requires the redshift
distribution of sources at all fluxes of interest.  In practice, we know
the number counts, the integral of the RLF over source redshift at fixed flux, over a very wide
range of fluxes from $10$ $\mu$Jy to $10^2$ Jy with reasonable accuracy, and 
the redshift distributions of bright sources ($S_5 > 0.3$ Jy).
There are weak, and systematically suspect, constraints on the local
RLF of fainter radio sources.   Because of the limitations from these data,
the redshift distribution of sources fainter than $0.3$ Jy is  largely 
determined by assumptions about the structure of the RLF such as smoothness
and evolution.  In this section we develop a method that will allow us to
explore the effects of these uncertainties on the expected number of lenses
in the flat spectrum samples.  There are, of course, many studies of the
RLF made with the goal of understanding the evolution of radio sources in
luminosity and density (e.g. Condon 1989, Peacock 1985, Kellerman \& Wall 1987, 
Windhorst et al. 1990, Dunlop \& Peacock 1990, Dunlop 1994).  
Generally, these RLF models are not well 
designed for studies of gravitational lenses because of their 
assumptions about unmeasured redshifts in redshift surveys of radio
sources.  We use the Dunlop \& Peacock (1990) pure-luminosity evolution 
model as our point of comparison to these earlier works.  

\subsection{Sources of Data}

We build the RLF from three elements: number counts as a function of flux, 
estimates of the local RLF, and limited redshift surveys.
We collated the 5~GHz number counts from Altschuler 
(1986), Bennett et al. (1985),
Donnelly et al. (1987), Fomalont et al. (1984), Fomalont et al. (1991), 
Gregory \& Condon (1981), Maslowski et al. (1981), Pauliny-Toth et al. (1978), 
and Wrobel \& Krause (1990) and the 5~GHz flat spectrum number
counts from Condon \& Ledden (1981), Donnelly et al. (1987), Fomalont et al. 
(1984), Fomalont et al. (1991), Owen et al. (1983), Pauliny-Toth et al. (1978) 
and Witzel et al. (1979).  The data for the flat spectrum counts are more 
limited than the overall number counts, so we fit the fraction of the sources 
that have a flat spectrum as a function of flux $S$ with a low order 
polynomial in $\log S$ and then correct the overall counts
using the estimated fraction.  The errors in the differential number counts were
broadened to include the statistical and fit errors for the fraction of
flat spectrum sources.  We use this method because the fraction of flat spectrum
sources is only measured in averages over large flux ranges, but varies slowly
with flux (it ranges from $\sim 0.25$ to $\sim 0.5$).  We avoid over-smoothing 
the counts by fitting the fraction of sources that are flat and then 
correcting the total counts.

The local RLF is determined from a combination of the complete redshift
surveys of bright radio sources and radio surveys of optical magnitude
limited samples of nearby galaxies.  The latter surveys constrain the
local RLF in the observed flux range from $\sim 10$ mJy to $\sim 500$ mJy.
We use the models of Toffolatti et al. (1987) and 
Dunlop \& Peacock (1990).  The faint local RLF data have large statistical
and systematic errors as there are few nearby flat-spectrum sources, and
the data are derived from heterogeneous optical magnitude and galaxy-type
limited surveys.  We tried models without the local RLF constraints, 
and these models show significantly larger variations in the expected number of lenses.

There are nearly complete redshift surveys of flat spectrum radio sources 
for fluxes $S_{2.7} > 1.5$ Jy from Peacock \& Wall (1981) and
Wall \& Peacock (1985).  We assumed a spectral index of $\alpha=0$
for the compact sources ($S_\nu \propto \nu^{-\alpha}$) to convert the 2.7~GHz 
fluxes to 5~GHz  following Dunlop \& Peacock (1990). We divided the data 
into three samples for the flux ranges from $3\hjy < S_{2.7} < 10\hjy$ (34 
sources, 33 redshifts), $2\hjy < S_{2.7} < 3\hjy$ (37 sources, 36 redshifts), 
and $1.5\hjy < S_{2.7} < 2\hjy$ (31 sources, 29 redshifts).   The missing 
redshifts in these bright source surveys are usually BL Lac objects with 
featureless spectra. The number of sources appears small because we include 
only the flat-spectrum subsets of larger surveys. For fainter sources the 
redshift surveys are very incomplete.  From Peacock (1985) we 
have a sample with $0.5 \hjy < S_{2.7} < 1.5 \hjy$ (40 sources, 30 redshifts),
and from the Parkes Selected Area Survey (PSAS, Dunlop et al. 1986,
Dunlop et al. 1989, Allington-Smith, Peacock
\& Dunlop 1991) we have a 
sample with $0.1 \hjy < S_{2.7} < 0.5 \hjy$ (34 sources, 21 redshifts).  
The CJI survey (Polatidis et al. (1995), Thakkar et al. 
(1995), Xu et al. (1995)) has the flux range $0.7 \hjy < S_5 < 1.3 \hjy$ (73 sources, 59 
redshifts), and the CJII survey (Taylor et al. 1994, 
Henstock et al. 1995, Henstock, Browne, \& Wilkinson 1994) has the flux range 
$0.35 \hjy < S_5 < 0.7 \hjy$ (187 sources, 137 redshifts).  
The Parkes Half-Jansky Flat-Spectrum Sample (PHFS, Webster et al. 1995) 
with $ S_{2.7} > 0.5$ Jy (323 sources, 258 redshifts) is available only
as a redshift histogram for the entire sample.  We did not use the PHFS
sample because of its incompleteness, lack of flux information, and because
it probably has a large overlap with the other high flux samples.
For each survey where we could identify the objects, we used NED\footnote{The NASA/IPAC
Extragalactic Database (NED) is a project of the Jet Propulsion Laboratory, at the 
California Institute of Technology, under contract with NASA.}
to fill in any redshifts found after the original publication.

In both the number counts data and the redshift data there is some double 
counting of sources, so the samples from different studies are not fully 
statistically independent.  Such double counting gives extra weight to some 
measurements, but should not significantly bias our final results.  Moreover, 
we cannot eliminate double counting in the redshift distributions for the 
binned samples (like CJII) because we were unable to obtain the source and
redshift lists.

\subsection{Numerical Method}

Rather than fitting a parametric form for the luminosity function, we will
simply determine the two-dimensional luminosity function in redshift and flux 
$\rho=dN/dV d\ln S$ where $dV$ is the comoving volume element, and $S$ is the 
observed $5$~GHz flux (in Jy).  The luminosity of the source (W Hz$^{-1}$)
at the $5$~GHz {\it rest} 
frequency is $L=4\pi S D_{OS}^2 (1+z)^{1+\alpha}$ if the spectrum is 
$S_\nu \propto \nu^{-\alpha}$ and $(1+z)D_{OS}$ is the luminosity distance 
to the source.  We use flux rather than luminosity to describe the RLF to 
minimize the effects of cosmology and variations in the spectral index.

The RLF is specified on a grid in redshift and flux, where 
$\rho_{ij} = d N / d V d\ln S (z_i,S_j)$ is the mean comoving source density in the redshift range
$z_{i-1/2} < z < z_{i+1/2}$ and in the flux range $S_{j-1/2} < S < S_{j+1/2}$.
The flux zones are uniform in $\ln S$ with $j=1 \cdots N_S$ and $S_{1/2}=10$
$\mu$Jy and $S_{N_S+1/2}=10$ Jy with $N_S=101$.
The $j^{th}$ zone is centered at flux $S_j=(S_{j+1/2}S_{j-1/2})^{1/2}$.
Some physical constraints we try to impose
on the RLF model, such as a constant comoving density of sources, must compare
source densities at fixed luminosity density rather than fixed flux.  
We made a special choice for the redshift grid to simplify such calculations.  
The luminosity of a source with flux $S_j$ at redshift $z_i$ is
$L_{ij} = 4 \pi S_j D_{OS}^2(z_i) (1+z_i)^{1+\alpha}$.  We define the redshift zones
by the requirement that $L_{ij} = L_{i-1j+1}$, so that a line of constant
luminosity crosses the redshift-flux grid at 45$^\circ$ to the lines of constant flux.
This requirement leads to redshift zones that are roughly
logarithmic in $z_i$, with $z_1 =0.01$, $z_{N_z}=5.16$, and $N_z=110$ zones for $N_S=101$.  
The source density is assumed to be zero for higher redshifts, but we did not
force any regularity constraint at the redshift cutoff. 
The comoving volume element is $dV = 4\pi D_{OS}^2 d D_{OS}$ for flat 
cosmologies (see Carroll, Press, \& Turner 1992, eqn. (12)).  
Let the comoving volume
between redshift zero and $z_{i+1/2}$ be $V_{i+1/2}$, so that the comoving
volume element between redshifts $z_{i-1/2}$ and $z_{i+1/2}$ is
$\dv_i=V_{i+1/2}-V_{i-1/2}$.  The logarithmic flux volume element at flux
$S_j$ is $\df_j = \ln S_{j+1/2}/S_{j-1/2}$.  We use the numerical variable
$\alpha_{ij}$ with $\rho_{ij} = \exp(\alpha_{ij})$ to force the RLF to
be positive definite. 

The numerical solution must optimize the fit to the number counts as a 
function of flux, the local RLF, the redshift surveys of brighter sources,
the smoothness of the solution, and the number of lenses.
For each term related to the data we calculate an estimate of the likelihood
that the model fits the data, and a good model maximizes the likelihood of fitting the
data.  However, our model has many more degrees of freedom than there are constraints,
so we can find models that fit the data nearly perfectly. Such models also have odd 
wiggles and oscillations because they overfit the statistical fluctuations in the
data to achieve perfect agreement.  To balance this tendency we add smoothing functions
that are designed to be small when the solution is physically reasonable.  The smoothing terms
drive the solution to show small pixel-to-pixel variations, have larger numbers of faint than bright 
sources, and show varying amounts of evolution.  The procedure we use is linear regularization
(see Press et al. 1992).  We define $L_{data}$ to behave like a $\chi^2$ statistic, where 
a perfect fit has $L_{data}=0$ and a typical good solution has $L_{data}=N_{data}$ where $N_{data}$ 
is the number of constraints. Viewed as a maximum likelihood problem, we have defined $L_{data}$
to be $L_{data} = - 2\ln(\hbox{likelihood})$.
If $U$ is a concave smoothing function with minimum value $U=0$ for a perfectly smooth
solution, then our procedure adjusts the RLF to minimize the function
\begin{equation} 
   F = L_{data} + \lambda_U U
\end{equation}
for a fixed value of the Lagrangian multiplier $\lambda_U$ using the conjugate gradient
method (see Press et al. 1992).  If $\lambda_U=0$ we optimize
only the fit to the data, and we find a solution with $L_{data} \ll N_{data}$ that is
not very smooth. In the opposite limit, $\lambda_U \gg 1$, we optimize only the smoothness
of the solution to find $U \sim 0$ but $L_{data} \gg N_{data}$.  By varying $\lambda_U$,
we find the smoothest solution that is a statistically reasonable fit to the data
with $L_{data} = N_{data} \pm (2N_{data})^{1/2}$.  We search the space of reasonable 
RLFs consistent with the constraints by varying the structure of the smoothing 
function.  Physical results should not (and do not) depend on the
exact converged value of $L_{data}$. In this section
we describe the terms for fitting the number counts, the local RLF, 
the redshift surveys, and the smoothing.  In the next section we
describe the terms for the lensing data.

The number counts data consists of $k=1 \cdots N_N=61$ measurements of the average 
differential number counts $D_k$, where 
\begin{equation}
 D_k= \left\langle {  dN \over dS } \right\rangle = { 1 \over {S_{Hk} - S_{Lk} } }
    \int_0^\infty dz {d V \over d z} \int_{S_{Lk}}^{S_{Hk}} { d S \over S } \rho(z,S), 
\end{equation}
and $S_{Lk} < S < S_{Hk}$ is the flux range of the measurement.  The measurement 
uncertainty is $e_k$.  The current numerical model predicts that the average
differential flux counts are
\begin{equation}
    D_k^M = { 1 \over {S_{Hk} - S_{Lk} } }
              \sum_{i=1}^{N_z} \sum_{j=j_{min}}^{j_{max}} \rho_{ij} \dv_i
              \ln\left[ { \hbox{min}(S_{Hk},S_{j+1/2}) \over
                          \hbox{max}(S_{Lk},S_{j-1/2}) }
                 \right]
\end{equation}
where the limits of the flux summation are the flux zones bracketing the
constraint ($S_{j_{min}-1/2} < S_{Lk} < S_{j_{min}+1/2}$ and
$S_{j_{max}-1/2} < S_{Hk} < S_{j_{max}+1/2}$), and the limits of the redshift
summation cover all redshifts.  The numerical integral in eqn. (3) uses a 
$0^{th}$-order approximation, because it does no interpolation of the 
density values. Numerical
experiments showed that the grid resolution was high enough to avoid using
more complicated, higher order integration methods.  We estimate the likelihood 
of the model fitting the number counts data with a $\chi^2$ statistic,
\begin{equation}
  \chi^2_N = \sum_{k=1}^{N_N} \left( { D_k - D_k^M \over e_k } \right)^2,
\end{equation}
where we expect a good solution to have $\chi^2_N = N_N \pm (2N_N)^{1/2}$.  
The constraints extend from $10$ $\mu$Jy to $10$ Jy,
and we have no need to extrapolate to higher or lower fluxes than
are constrained by the number counts data.
The local RLF simply specifies the mean value of $\rho_{ij}$ over some flux
range to some low redshift limit (e.g. $0.07$ for Toffolatti et al. 1987).  
This leads to a $\chi^2_L$ for fitting the local RLF in which the
redshift summations of eqn. (3) extend only to the redshift limit of the 
local RLF.  Although we use 10 data in the
local RLF constraints, forcing $\chi^2_L=10$ represents an overfitting
of the local RLF data.  The RLF data should not be modeled using a $\chi^2$
distribution because the data for each point are derived from small numbers
of sources (0 to 5).  Broader Poisson uncertainties would be more appropriate,
but we were forced to use a $\chi^2$ statistic by the format of the published
local RLF models.
We also used both the Dunlop \& Peacock (1990) and Toffolatti et al.
(1987) models simultaneously to include the uncertainties from their differing
interpretations of the same data.

We fit the redshift surveys using binned data.  Binning the data is cruder 
than the two-dimensional Kolmogorov-Smirnov test developed and used by 
Peacock (1983) and Dunlop \& Peacock 
(1990), but adequate for our purposes.  Moreover, some of the redshift constraints
are available only in binned forms (the PHFS and CJII samples).
We divided the various samples used by Wall \& Peacock (1985) and Dunlop \&
Peacock (1990) into flux ranges and added the CJI and CJII samples 
to give a total of $k=1 \cdots 7$ redshift surveys whose properties
are summarized in \S2.1. Survey $k$ has $N_{Zk}$ measured redshifts 
and $N_{Ok}$ objects between fluxes
$S_{Lk} < S < S_{Hk}$.  The number of measured redshifts is generally less
than the number of objects, so the mean completeness of a survey is
$N_{Zk}/N_{Ok}$.  Survey $k$ is divided into $l=1 \cdots N_k$ redshift
bins bounded by $z_{Lkl} < z < z_{Hkl}$ (and $z_{Lkl}=z_{Hkl-1}$), and bin
$l$ of survey $k$ contains $N_{kl}$ objects with measured redshifts 
($\sum_{l=1}^{N_k} N_{kl} = N_{Zk}$).  To compare the model to the redshift
data we must normalize the model to match the number of objects of the
survey, and correct for incompleteness.

Let $A_k$ be the ``effective area'' of survey $k$, and let $f_{kij}$ be the
probability that a redshift is measurable for a source at redshift $z_i$ with
flux $S_j$ in survey $k$.  For simplicity we assumed that the redshift completeness 
in a given survey does not depend on the radio flux, so that $f_{kij} = f_{ki}$.  
The ``effective area'' is determined by the constraint that the total number of 
objects $N_{Ok}$ must match the flux counts,
\begin{equation}
   N_{Ok} = A_k \int_0^\infty dz {d V \over dz} 
                \int_{S_{Lk}}^{S_{Hk}} { dS \over S } \rho  
          = A_k \sum_{i=1}^{N_z} \sum_{j=j_{min}}^{j_{max}} \rho_{ij} \dv_i
              \ln\left[ { \hbox{min}(S_{Hk},S_{j+1/2}) \over
                          \hbox{max}(S_{Lk},S_{j-1/2}) }
                 \right].
\end{equation}
The number of objects with measured redshifts in bin $l$ is 
\begin{equation}
     N_{kl}^M = A_k \sum_{i=i_{min}}^{i_{max}} \sum_{j=j_{min}}^{j_{max}} f_{ki} \rho_{ij}
              \left[ \hbox{min}(V_{Hkl},V_{i+1/2}) -
                     \hbox{max}(V_{Lkl},V_{i-1/2}) \right]
              \ln\left[ { \hbox{min}(S_{Hk},S_{j+1/2}) \over
                          \hbox{max}(S_{Lk},S_{j-1/2}) }
                 \right]
\end{equation}
subject to the constraint that $N_{Zk} = \sum_{l=1}^{N_{k}} N_{kl}^M$.
$V_{Lkl}$ and $V_{Hkl}$ are the comoving volumes for redshifts less than
$z_{Lkl}$ and $z_{Hkl}$ respectively, and $i_{min}$ and $i_{max}$ are defined 
to be the zones bracketing redshift bin $l$ (i.e., 
$z_{i_{min}-1/2} < z_{Lkl} < z_{i_{min}+1/2}$).

Dunlop \& Peacock (1990) relied on redshift-magnitude relations, or simply 
assumed an intermediate redshift for the objects without measured redshifts.  
This is not adequate for a lensing calculation because the lensing 
probabilities vary strongly with redshift for $z>1$ where the scatter in 
the redshift-magnitude relations is too large to make them useful predictors.  
We instead parametrize our uncertainties through a completeness model defined 
by the $f_{ki}$, and plead for better redshift surveys in the conclusions.  

We treated two simple cases.  In the {\it uniform completeness} model (labeled 
by $C=0$ in Table 1), we assume that the completeness is independent of 
redshift, $f_{ki} = f_k$, and that the measured redshifts are an unbiased 
representation of the redshift distribution.  The probability of measuring 
any redshift is simply the average completeness of the sample,  
$f_k = N_{Zk}/N_{Ok}$.  In the {\it linear completeness model} 
(labeled by $C=1$ in Table 1), $f_{ki} = a + b z_i$ with $b < 0$ so that
the completeness declines with redshift.  If $\langle z_k \rangle$ is the mean 
redshift predicted by the model, then we use $a=1$ and 
$b= -(N_{Ok}-N_{Zk})/(N_{Ok}\langle z_k \rangle)$.  The model for $f_{ki}$
must be modified if the detectable fraction becomes negative at
the maximum redshift, $f_{kN_z} < 0$ at  $z_{max}=z_{N_z+1/2}$.  When
it becomes negative, we must drop the assumption that the completeness
is unity at zero redshift ($a=1$), and instead use the coefficients
$a=(N_{Zk}/N_{Ok}) z_{max}/(z_{max}-\langle z \rangle)$ and $b=-a/z_{max}$.  
The two completeness models have negligible differences in nearly complete 
surveys, but the
linear completeness model will predict larger numbers of high redshift sources
in incomplete surveys.

For a given completeness model, we compare the predicted number of measured
redshifts $N_{kl}^M$ to the measured number $N_{kl}$. We do this using the logarithm
of the Poisson maximum likelihood ratio for that point,
\begin{equation} 
   L_{kl} =  N_{kl} \ln { N_{kl}^M \over N_{kl} } - N_{kl}^M + N_{kl} 
\end{equation}
so that when $N_{kl}^M=N_{kl}$, $L_{kl}=0$.
If we take the weighted average of $L_{kl}$ over the Poisson
likelihood function, we find that the mean value of the likelihood is 
$\langle L_{kl}\rangle=-0.5$ for $N_{kl} >> 1$.  For $N_{kl}=0$, $1$, and $2$, 
we find $\langle L_{kl} \rangle = -1$, $-0.58$, and $-0.54$ respectively.   
The total likelihood of fitting redshift survey $k$ is 
\begin{equation}
  L_k =  \sum_{l=1}^{N_k} L_{kl}.
\end{equation}
The function $L_{data}$ is equal to $-2\ln (\hbox{likelihood})$, so to give
equal statistical weight to the redshift data and the number counts
we add $-2 L_k$ to $L_{data}$.

The final terms of the function $F$ (eqn. (1)) are designed to produce a smooth
model.  We must add such terms because the number counts, redshift, and lens
constraints are not sufficient to determine fully the density of sources at
all redshifts and fluxes -- the data are noisy, and redshift surveys are available only
for bright sources ($ S \gtorder 300$ mJy).
By construction the solution is positive definite and/or monotonic depending
on whether we use $\alpha_{ij}$ or $\beta_{ij}$ as the numerical variable.
We introduce terms to bias the solution to a smoothly increasing number 
of sources at fainter fluxes and either no evolution or strong evolution
at fixed luminosity.  The first smoothing term biases the solution to a slope 
of $\rho \propto S^{-\xi}$ at a fixed redshift, by adding the function
\begin{equation}
      U_1 = \sum_{i=1}^{N_z} \sum_{j=1}^{N_f-1}
            \left[\alpha_{ij}-\alpha_{ij+1}-\xi \ln (S_{j+1}/S_j)\right]^2
\end{equation}
where our standard model uses the ``Euclidean slope'' of  $\xi=1.5$.  
The second term biases the solution to a constant comoving luminosity function
by adding
\begin{equation}
  U_2 =  \sum_{i=2}^{N_z} \sum_{j=1}^{N_S-1}  { 1 \over 1 + \kappa w_{i-1/2} }
            \left[\alpha_{ij}-\alpha_{i-1j+1}\right]^2.
\end{equation}
The third term biases the solution to constant comoving density at
fixed flux rather than fixed luminosity
\begin{equation}
  U_3 =  \sum_{i=2}^{N_z} \sum_{j=1}^{N_S} { \kappa w_{i-1/2} \over  1 +\kappa w_{i-1/2} }
            \left[\alpha_{ij}-\alpha_{i-1j}\right]^2.
\end{equation}
where $w_{i-1/2}$ is the ratio of the difference in time between redshifts 
$z_i$ and $z_{i-1}$ in units of the local Hubble time to the difference in 
time in units of the local Hubble time for redshift zones $N_z$ and $N_z-1$, 
and $\kappa$ is a constant.  We must introduce some weighting factor
$w_{i+1/2}$ because the redshift zones are (approximately) logarithmically
spaced, and uniform weighting allows too much evolution at low redshift.  
At redshift $z$, 
$w(z)^{-1} \simeq  1 + 1/3((1+z)^{1/2}-1)$, and
$\kappa w(z) =1$ at $z_c\simeq (6\kappa-5)/9(\kappa-1)^2$.  The term $U_2$ 
controls the smoothing for $z\ltorder z_c$, and the term $U_3$ controls the 
smoothing for $z\gtorder z_c$.  The larger $\kappa$ becomes, the stronger the 
bias towards luminosity evolution at lower redshifts.  These terms do not 
represent all possible, physical smoothing terms, and in \S5 we explore 
treating the lensing terms introduced in \S3 as a type of regularizing term.

In summary, the constraint term appearing in the function $F$ is 
$L_{data} = \chi_N^2 + \chi_L^2 - 2 \sum_{k=1}^7 L_k$ and the smoothing
term is $U = U_1 + U_2 + U_3$. There are a total of 147 constraints, so
we find the smoothest solutions with $L_{data}=147$ for different combinations
of the constants in the smoothing terms ($\xi$ and $\kappa$) and for the two
different completeness models.  We made one simplification to reduce the amount
of calculation.  {\it The smoothing terms and all the constraints on the RLF except the lensing
constraints are calculated in a fixed $\Omega_0=1$ 
Einstein-DeSitter cosmology. } For the data constraints this has no effect
because they only depend on the number per unit redshift and flux 
($dN/dzdS=\rho d V/dz$, which is cosmology independent), but the 
physical meaning of the smoothing constraints varies
with cosmology because $U_2$ and $U_3$ depend on the luminosity density of the sources.

\section{The Lens Model}

We use a (density) ellipsoidal singular isothermal sphere for our lens model,
characterized by a circular critical radius $b=4\pi(\sigma/c)^2 D_{LS}/D_{OS}$
and an ellipsoidal parameter $\epsilon$ (see Kassiola \& Kovner 1993).  The surface 
density axis ratio of the model is $b/a=(1-\epsilon)^{1/2}/(1+\epsilon)^{1/2}$. 
The circular critical radius $b$ depends on the (dark-matter) velocity 
dispersion $\sigma$ and the ratio of the distances from the lens to the 
source and from the observer to the source.  In a flat cosmology 
($\Omega_0+\lambda_0=1$), the distances are defined by 
\begin{equation}
  D_{12} = { c \over H_0} \int_{z_1}^{z_2} 
         dz \left[ (1+z)^2(1+\Omega_0 z) - z(2+z)\lambda_0 \right]^{-1/2} 
\end{equation}
(Carroll, Press, \& Turner 1992) where $\Omega_0$ and $\lambda_0$ are the
matter density and the cosmological constant.  If the lower redshift $z_1=0$,
these are proper motion distances.  The lens equations are
\begin{eqnarray}
    u &= &x - { b \over \sqrt{2\epsilon} } \tan^{-1} \left[
      { \sqrt{2\epsilon} \cos\theta \over (1-\epsilon\cos2\theta)^{1/2} }
         \right] \\
    v &= &y - { b \over \sqrt{2\epsilon} } \tanh^{-1} \left[
      { \sqrt{2\epsilon} \sin\theta \over (1-\epsilon\cos2\theta)^{1/2} }
         \right] 
\end{eqnarray}
where $u,v$ are the source coordinates, $x,y$ are Cartesian
image coordinates, and $r,\theta$ are polar image coordinates.  
The magnification has a remarkably simple form, 
\begin{equation}
   M^{-1} = 1 - { b \over r } { 1 \over (1-\epsilon\cos 2\theta)^{1/2}},
\end{equation}
and contours of constant magnification are contours of constant surface
mass density.  In particular, the axis ratio of the tangential
critical line ($M^{-1}=0$) is the axis ratio of the density.

The lens produces both two- and four-image systems.  The cross sections are not 
expressible in simple analytic forms, although they are easily calculated 
numerically, and we can express the four-image, two-image and total cross 
sections as power series in $\epsilon$, with
\begin{eqnarray}
  \sigma_4 &= &{\pi \over 6 } \epsilon^2 b^2 \left[ 1 + 
      { 21 \over 20} \epsilon^2 + { 57 \over 56 } \epsilon^4 +
      { 187 \over 192 } \epsilon^6 \cdots \right] \\
  \sigma_2 &= &\sigma_t - \sigma_4 \quad\hbox{and} \\
  \sigma_t &= &\pi b^2 \left[ 1 + { 1 \over 3 } \epsilon^2 +
      { 1 \over 5 } \epsilon^4 + { 1\over 7 } \epsilon^6 \cdots \right].
\end{eqnarray}
These expressions are accurate to 5\% or better
for $b/a \ltorder 0.4$ ($\epsilon \ltorder 0.7$), and they are invalid
beyond $\epsilon > 0.73097$  where the astroid caustic first pierces
the radial (pseudo) caustic.  

From the properties of the tangential critical line (where $M^{-1}=0$) we can 
determine the asymptotic cross sections as a function of magnification 
(see Blandford \& Narayan 1986).  The asymptotic cross section for four-image 
systems with magnification greater than $M$ is 
\begin{equation}
 \sigma_4 (>M) = { 4\pi b^2 \over (1-\epsilon^2)^{1/2} } { 1 \over M^2 }
 \quad\hbox{for}\quad M >> 1
\end{equation}
and the asymptotic cross section for two-image systems with magnification
greater than $M$ is
\begin{equation}
  \sigma_2 (>M) = { 16 b^2 \over 5 \sqrt{3\epsilon} } 
     \left[ (1+\epsilon)^{-1/2} + (1-\epsilon)^{-1/2} \right] 
     { 1 \over M^{5/2} }
 \quad\hbox{for}\quad M >> 1
\end{equation}
and these expressions are valid for $\epsilon < 0.73097$.  

We calculated the integral cross sections $\sigma_2(>M,r)$ and 
$\sigma_4(>M,r)$ numerically for total magnifications greater than $M$ 
with flux 
ratios between the brightest and faintest images smaller than $r=30$.
Figure 1 shows the cross sections for galaxies with axis ratios
of $b/a=0.7$ and $b/a=0.65$ in units of the total cross
section of the equivalent circular lens, $\pi b^2$.  There is, of
course, a distribution of galaxy ellipticities (see Ryden (1992) or
Schechter (1987) for examples).  We are primarily
interested in the distribution for E and S0 galaxies because spirals
produce only 15\%-20\% of lenses (see Kochanek 1996).
Moreover, while it is dubious that the ellipticity of the
light in E and S0 galaxies is the true ellipticity of the total
mass distribution, it is certainly incorrect to assume so for spirals 
(see the review by Sackett 1996).  For simplicity, we will use the average 
cross sections for the axis ratio range $b/a=1$ to $b/a=0.5$ as our
basic model, and the integral cross sections for this model are 
also shown in Figure 1.   The average four-image cross section is
comparable to that of a galaxy with $b/a=0.65$ even though the average 
galaxy has only $b/a=0.75$ because the four-image cross section is
$\propto \epsilon^2$.  For this treatment we ignore the question
of whether the normalization of the potential $b$ depends on the
ellipticity, and in \S6 we examine the issue of the mean ellipticity
required to fit lens data in more detail.

\begin{figure}
\centerline{\psfig{figure=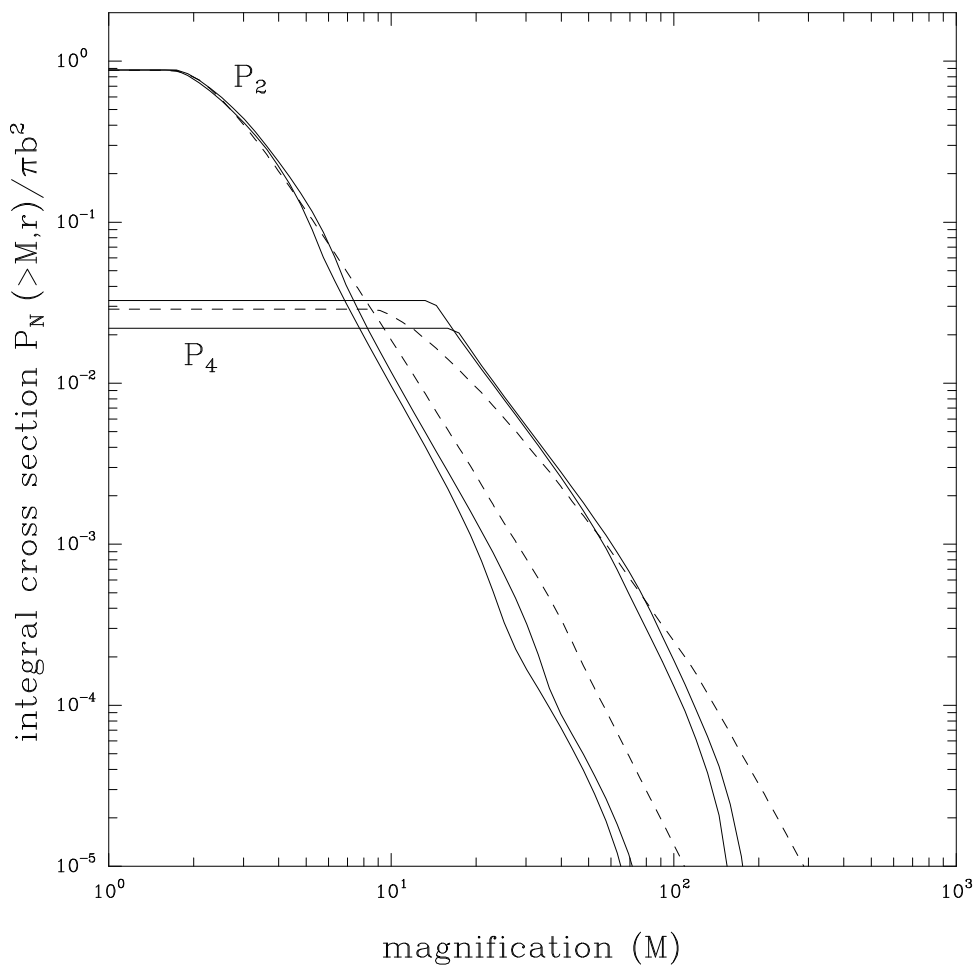,height=5.0in}}
\caption{Integral cross sections for two- and four-image systems
with flux ratios smaller than $r=30$.  The solid lines show
the cross sections for $b/a=0.70$ and $b/a=0.65$ ellipsoidal isothermal
spheres, and the dashed lines show the mean for a uniform
distribution of galaxies $0.50 < b/a < 1$, all in units of
the cross section for a circular lens $\pi b^2$.}
\end{figure}

Let $\hat{P}_2(>M,r)=\sigma_2(>M,r)/\pi b^2$ and 
    $\hat{P}_4(>M,r)=\sigma_4(>M,r)/\pi b^2$ 
be the probability that a two- or
four-image system has total magnification greater than $M$ and
a brightest to faintest image flux ratio smaller than $r$
normalized by the cross section of the equivalent circular
lens.  A convenient property of the singular ellipsoidal isothermal lens is
that the structure of the optical depth is unchanged from the singular
isothermal sphere.  We assume a selection function that
detects all images with flux ratios smaller than $r$ in the
separation range $\theta_{min} < 2 b < \theta_{max}$.
We assume a Schechter (1976) function exponent of $\alpha=-1$ and
a Faber-Jackson (1976) exponent of $\gamma=4$ to describe the number
counts of galaxies and the relation between luminosity and the velocity
dispersion of the isothermal sphere,
\begin{equation}
 { dn \over dL} = { n_* \over L_* } \left[ { L \over L_* } \right]^\alpha \exp(-L/L_*)
  \quad\hbox{and}\quad
  { L \over L_* } = \left[ { \sigma \over \sigma_*} \right]^\gamma,
\end{equation}
where $n_* = (0.61\pm0.21)h^3 10^{-2}$ Mpc$^{-3}$ is the local comoving density of E and 
S0 galaxies (Loveday et al. 1992, Marzke et al. 1994), and
$\sigma_* = (220\pm20) \kms$ is the (dark-matter) velocity dispersion of an
$L_*$ galaxy (Kochanek 1993, 1994, Breimer \& Sanders 1993, Franx 1993).  
The fraction of the lenses that have critical radii in the
detectable range is
\begin{equation}
      F_a =  30 \int_0^1 dx \, x^2 (1-x)^2 \left[
\exp\left( - \Delta\theta_{min}^2/\Delta\theta_*^2 (1-x)^2 \right) -
\exp\left( - \Delta\theta_{max}^2/\Delta\theta_*^2 (1-x)^2 \right) 
      \right]
\end{equation}
where the characteristic image separation is 
$\Delta\theta_* = 8\pi(\sigma_*/c)^2=2\parcs8 (\sigma_*/220\kms)^2$, and 
the mean image separation in flat cosmologies is $\Delta\theta_*/2$.
The fraction of lenses in the separation range from
$\theta_{min}=0\parcs3$ to $\theta_{max}=5\parcs0$ is $F_a =0.91$, and
the range of separations is large enough to make the value insensitive to
large changes in the limits.  For $\theta_{min}=0\parcs2$, the fraction
rises to $F_a=0.96$, for $\theta_{min}=0\parcs4$ it drops to $F_a=0.87$, 
for $\theta_{max}=3\parcs0$ it drops to $F_a=0.88$, and for 
$\theta_{max}=7\parcs0$ it rises to $F_a=0.92$.  The optical depth to 
lensing for circular isothermal spheres in flat cosmologies is 
$\tau = \tau_* (D_{OS}/r_H)^3/30$ where $r_H=c/H_0$ (Turner 1990). 
The optical depth scale is
$\tau_* = 16\pi^3 n_* r_H^3 (\sigma_*/c)^4 \Gamma[1+\alpha+4/\gamma]= 0.024\pm0.012$,
where the uncertainties are dominated by $n_*$ and $\sigma_*$.

We want to determine the number of lenses in a survey of $N$ sources brighter
than a flux limit $S_0$.  As with the redshift surveys, we must first determine
the effective area of the survey $A_L$ to normalize the number of sources 
being examined, 
\begin{equation}
   N = A_L \int_0^\infty dz {d V \over dz} 
                \int_{S_0}^\infty { dS \over S } \rho(z,S)
       = A_L \sum_{i=1}^{N_z} \sum_{j=j_{min}}^{N_S} \rho_{ij} \dv_i
              \ln\left[ { S_{j+1/2} \over
                          \hbox{max}(S_0,S_{j-1/2}) }
                 \right]
\end{equation}  
where $S_{j_{min}-1/2} < S_0 < S_{j_{min}+1/2}$ brackets the flux limit of the 
survey.  At redshift $z_i$ and flux $S_j$, the fraction of the sources that 
are lensed into the survey is just the optical depth $\tau(z_i)$, multiplied 
by the fraction of lenses with detectable separations $F_a$, multiplied by 
the probability that the source is sufficiently magnified to have a flux 
larger than the survey flux limit $P_n(>S_0/S_j,r)$, where $n=2$ or $4$ for the 
two-image and four-image lenses respectively.  Thus, the expected number
of $n$ image lenses is just
\begin{equation}
   N^M_n = A_L \sum_{i=1}^{N_z} \sum_{j=1}^{N_S} \rho_{ij} \dv_i \df_j
                     \tau(z_i) F_a P_n(> S_0/S_j,r)
\end{equation}  
including the selection effects on both the separation and the flux ratios.  
The total number of lenses is $N_T^M=N_2^M+N_4^M$.  {\it Remember that in 
equations (23) and (24), the optical depth $\tau(z_i)$ is computed in the 
current cosmology, but the two parts of the cosmology independent term 
$\rho dV/dz$ are computed in an $\Omega_0=1$ model.} Simple modifications
of eqn. (24) can be used to determine the distribution of the lenses in source 
redshift or flux.  Because of the increased complexity and the uncertainties 
arising from fitting the RLF, we chose not to include the uncertainties in the 
lens models or the more complicated configuration probabilities used in 
Kochanek (1993, 1996). 
 
We define the likelihood of fitting the lens data by the logarithm of the Poisson likelihood 
ratio (as in the redshift survey models of \S2.2), with
\begin{equation} 
  L_L = N_T \ln { N_T^M \over N_T } - N_T^M + N_T 
\end{equation}
rather than using two separate terms for the numbers of two- and four-image lenses.
The total number of lenses depends only weakly on the ellipticity, so it is a more
robust variable for exploring the cosmological implications of the radio surveys
in \S4 and \S5.  We add this likelihood to the overall function with an additional Lagrangian
multiplier $\lambda_L$ that will determine the weight given to the lensing constraint
in fitting the RLF,
\begin{equation}
    F = \chi_S^2 + \chi_L^2 - 2 \sum_{k=1}^7 L_k  
       + \lambda_U (U_1+U_2+U_3)
      - 2 \lambda_L L_L .
\end{equation}

\section{Models Without Lensing Constraints}

In this section we consider models constrained to fit only the number counts
and redshift data ($\lambda_L=0$ in eqn. 26).  We start from the Dunlop \& Peacock (1990)
pure luminosity evolution model, and then adjust $\lambda_U$ until we have
a ``typical'' good fit to the data with $L_{data} = 147$.  We have several free 
parameters to explore in this process: the redshift completeness model, the bias exponent
$\xi$ for smoothing term $U_1$, the balance $\kappa$ between 
$U_2$ and $U_3$, and the background cosmological model.  We examine only the
total number of lenses in this section, and defer a discussion of the
relative numbers of two- and four-image lenses to \S6.  We do not discuss the
form of the RLF {\it per se}.

\begin{deluxetable}{rrrrrrrrrrr}
\tablenum{1}
\tablecaption{Results With No Lensing Constraints}
\startdata
Model  &$\Omega_0$ &C &$\xi$ &$\kappa$ &\multicolumn{2}{c}{JVAS} &\multicolumn{2}{c}{PHFS}
                                       &\multicolumn{2}{c}{CLASS} \nl
       &           &  &      &         &$N_2$ &$N_4$ &$N_2$ &$N_4$ &$N_2$ &$N_4$ \nl
00     &1.0        &0 &--    &--       &2.44  &0.48  &0.36  &0.09  &2.71  &0.44  \nl 
01     &1.0        &0 &1.5   &0        &2.22  &0.40  &0.37  &0.09  &2.29  &0.30  \nl
02     &1.0        &1 &1.5   &0        &2.61  &0.45  &0.45  &0.10  &2.60  &0.32  \nl
03     &1.0        &1 &2.0   &0        &2.78  &0.49  &0.47  &0.11  &2.80  &0.34  \nl
04     &1.0        &1 &1.0   &0        &2.82  &0.49  &0.47  &0.11  &2.84  &0.35  \nl
05     &1.0        &1 &1.5   &1        &3.28  &0.61  &0.54  &0.13  &3.46  &0.46  \nl
06     &1.0        &1 &1.5   &2        &3.21  &0.58  &0.54  &0.13  &3.33  &0.44  \nl
07     &1.0        &0 &1.5   &1        &2.76  &0.53  &0.45  &0.11  &3.00  &0.42  \nl
08     &1.0        &0 &1.5   &2        &2.83  &0.54  &0.47  &0.12  &3.06  &0.43  \nl
09     &1.0        &0 &1.5   &3        &2.60  &0.49  &0.44  &0.10  &2.75  &0.38  \nl
\enddata
\tablecomments{
$C=0$ is the uniform redshift completeness model, and $C=1$ is the linear redshift 
completeness model.  The slope $\xi$ appears in the smoothing term $U_1$ (eqn. 9), and 
$\kappa$ determines the balance between $U_2$ (eqn. 10) and $U_3$ (eqn. 11).  
Larger values of $\kappa$ drive the solution to greater luminosity evolution.  
$N_2$ and $N_4$ are the expected numbers of two- and four-image systems with 
flux ratios smaller than $r=30$, uncorrected for detectable separations 
(multiply by $F_a=0.91$) or extended sources (multiply by $0.95$).}
\end{deluxetable}

%

Table 1 explores models consistent with the number counts and redshift data
under various assumptions about the completeness and the smoothing parameters 
in an $\Omega_0=1$ cosmological model.  All the models are converged to fit 
the data equally well, and no constraint is fit poorly.  The initial Dunlop 
\& Peacock (1990) model, labeled 00, is a reasonably good fit to both 
the number counts and redshift data.  It has a value of $L_{data}=209$, while
our target value is $L_{data}=147$, so the typical constraint is fit within 
``1.5$\sigma$''.  The largest discrepancies are in the faint number counts
(too few faint sources), and the $0.5$ to $1.5$ Jy redshift 
distribution (too few high redshift sources).  If the observed redshifts
are a fair sample (completeness model 0) or we bias the solution towards no
luminosity evolution ($\kappa=0$), then we find smaller numbers of lenses  
than in the models with decreasing sample completeness with redshift, or
strong luminosity evolution.  The slope of the bias function at fixed redshift,
$\xi$, only weakly affects the results.  The total number of lenses predicted 
for the JVAS sample ranges from 2.6 (Model 01) to 3.9 (Model 05), and 
14--16\% of the total are four-image lenses.  For comparison, the Dunlop \& 
Peacock (1990) Model 00 predicts 2.9 lenses, 16\% of which are four-image 
lenses, as also found by King \& Browne (1996).
The spread in the expected number of lenses in Table 1 is a factor of 3/2,
comparable to the uncertainties from the other components of the lens 
model (see \S3).  Presumably the true range is somewhat broader because
we explored only a finite number of smoothing terms -- we return to this
issue in \S5.  

The numbers in Table 1 were not corrected for the limited range of detectable
separations, which we estimated would reduce the number of lenses detected
by the factor $F_a=0.91$ in \S3.  Our statistical model is for point
sources, so we should exclude the 5\% of JVAS sources that have extended
structure (see Patnaik et al. 1992), and treat the statistics of the 95\% of the 
sources that are unresolved or point-like.  The statistics of sources with 
extended or multiple structure are strongly modified by the effects of the 
structure (see Kochanek \& Lawrence 1990), and the probability
of an object in this subpopulation being lensed can be much
larger than for a point source of comparable flux.
If we drop the extended sources, we must also drop any lenses produced from
this population.  In the JVAS survey, this means eliminating B 1938+666,
a lensed double with at least one extended component.
Wherever we quantitatively compare the survey data to the observed numbers
of lenses, we include the angular selection factor and eliminate the
extended sources.  In adopting this simple approach, we assume 
that the extended sources are otherwise indistinguishable in their
redshift and flux distributions from the compact sources.  With these
corrections, the observed sample contains 4 lenses while we expected to find
2.27 (3.37) lenses in Model 01 (05).  The Poisson probabilities
are 11\% and 18\% respectively, compared to a peak Poisson
probability of 19.5\%.  In short, the numbers of lenses found in the JVAS
survey are statistically compatible with the best fit lensing model found
for the quasar lens surveys and an $\Omega_0=1$ cosmological model.

Figure 2 shows the expected number of lenses for Models 01 and 05 in the JVAS, 
CLASS, and PHFS surveys as a function of the matter density $\Omega_0$ in cosmologies
with a cosmological constant, $\Omega_0+\lambda_0=1$.
In these models we fixed the redshift and flux distribution of the sources
(i.e. $dN/d S d z = (dN/dSdV)(dV/dz)$) not the comoving luminosity function
($dN/dLdV$).  This simplification means that the effective smoothing
function changes with the cosmological model, but otherwise does not
affect the results.
Including the angular selection function and dropping the
extended sources, the maximum likelihood cosmological model for the JVAS
sample ranges from $\Omega_0\simeq 0.5$
for Model 01 to $\Omega_0\simeq 0.85$ for Model 05.  The 90\% (99\%) confidence
lower limits are $\Omega_0 \gtorder 0.2$ ($\gtorder 0.15$) for Model 01,
and $\Omega_0 \gtorder 0.4$ ($\gtorder 0.3$) for Model 05.  The upper limits
are not as well defined.  Model 01 has a 90\% confidence upper limit of
$\Omega_0 \ltorder 1.5$,  and Model 05 has a 90\% confidence
limit of $\Omega_0 \ltorder 2.0$.  These limits are consistent with those found
from the statistics of lensed (optical) quasars (see Kochanek 1996).  Note
that the Poisson uncertainties in the cosmological model are comparable
to the systematic uncertainties from the structure of the luminosity function.
Clearer cosmological limits from the radio surveys will depend on both 
finding more lenses and more tightly constraining the luminosity function.

Figure 3 shows the expected number of lenses per $10^3$ sources
brighter than a given flux in several of the models, and the error bars or 
limits that can be derived for each survey.
The number of lenses expected in the CLASS survey is always close to that 
expected for the JVAS survey.  Although the total number of sources in 
the CLASS survey is larger (3258 versus 2200 sources), they are fainter
(a flux limit of 25--50 mJy versus 200 mJy).  The weaker magnification
bias balances the larger numbers of sources.  The PHFS survey is smaller
(323 sources) but brighter (0.5 Jy) than the JVAS survey.  Although the
magnification bias is somewhat larger than in the JVAS survey, the number
of sources is so much smaller that we typically expect only 0.3--0.6 lenses in
the PHFS survey.  The flattening probability of finding four-image lenses 
near 10 mJy is caused by the steepening of the number counts near 0.1-10 mJy.

\begin{figure}
\centerline{\psfig{figure=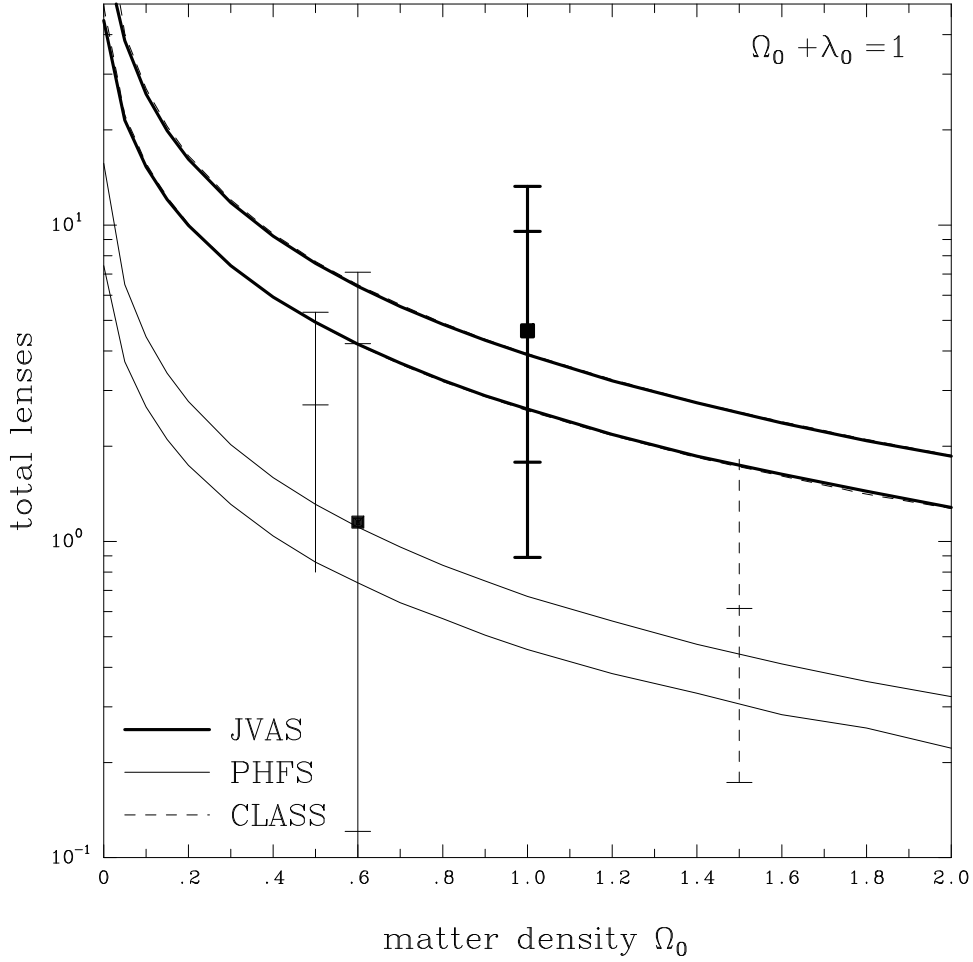,height=5.0in}}
\caption{The expected numbers of lenses in the JVAS (heavy solid line), CLASS 
(dashed line, under heavy solid), and PHFS (light solid line) flat spectrum lens surveys as a 
function of the matter density $\Omega_0$ in flat cosmological models with a 
cosmological constant ($\Omega_0+\lambda_0=1$).  The upper (lower) line of each pair 
is for Model 05 (Model 01).  The JVAS error bar (heavy solid) shows the maximum 
likelihood value, and the 90\% and 99\% confidence intervals.  The CLASS error bar 
(dashed) shows where the Poisson probability of 
finding 2 or more lenses exceeds 1\% (lower bar), 10\% (upper bar), and 50\% 
(upper tip of line).  The two PHFS error bars show either the 
maximum likelihood limits if the survey contains one lens, or the 1\% 
(upper bar), 10\% (lower bar), and 50\% (lower tip of line) Poisson
limits for finding 
no lenses. The limits for the surveys are corrected for separations and do not 
include extended sources. There is no significance to the location of the error
bars in $\Omega_0$.}
\end{figure}

\begin{figure}
\centerline{\psfig{figure=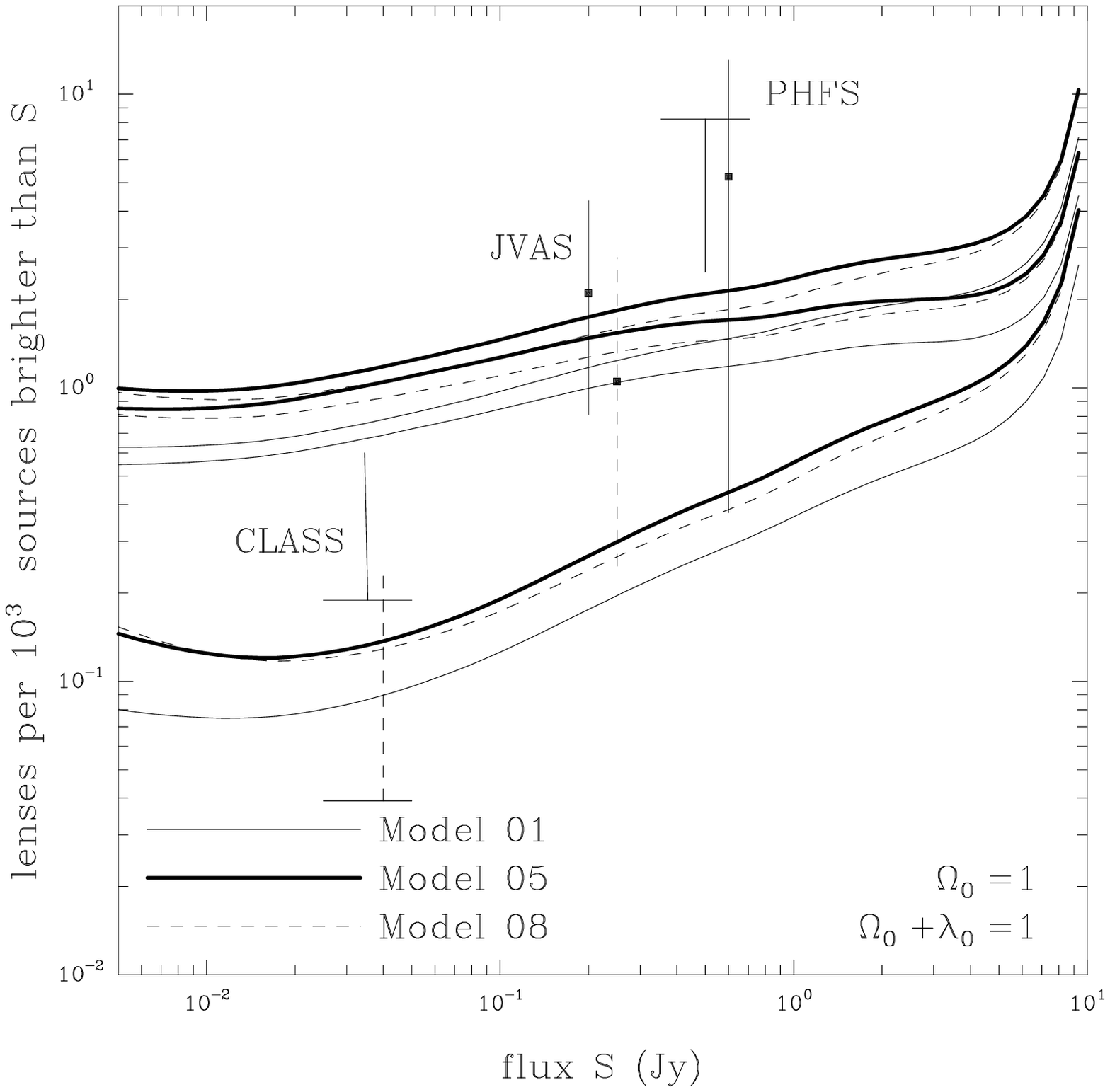,height=3.5in}
            \qquad
            \psfig{figure=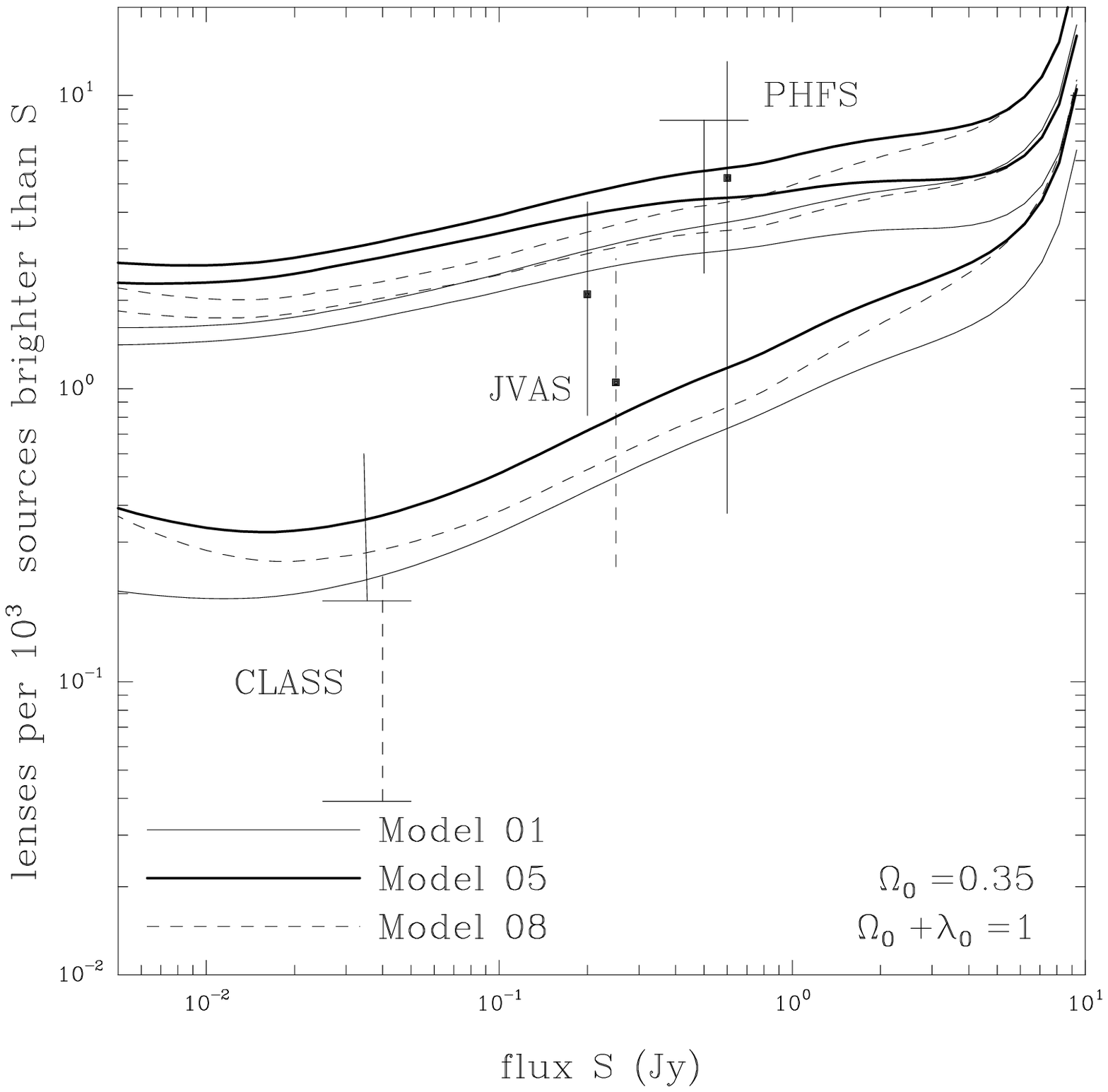,height=3.5in}}
\caption{The expected number of lenses per $10^3$ sources brighter than flux limit 
 $S$ for Models 01 (light solid), 05 (heavy solid), and 08 (dashed) in
 either an $\Omega_0=1$ (left) or an $\Omega_0=0.35$ (right) flat cosmological model.
 Three curves are shown for each model: the top curve is the total number of 
 lenses, the middle curve is the number of two-image lenses, and the lower 
 curve is the number of four-image lenses.  For the JVAS survey ($S=0.2$ Jy) 
 the maximum likelihood (filled point) and 90\% confidence 
 limits are shown on the total number of lenses (solid) and the number 
 of four-image lenses (dashed, offset).  For the CLASS survey ($S=25$ to 
 $S=50$ mJy) the horizontal line marks the limit for a 10\% Poisson 
 probability of finding at least two lenses (solid) or at least one four-image 
 lens (dashed, offset).  The vertical lines extend to the point
 where there is a 50\% Poisson probability.  For the PHFS sample ($S=0.5$ Jy), 
 the upper limit (left) marks the point where there is a 10\% probability of 
 finding no lenses in the sample, and the line extends to the point where 
 there is a 50\% probability of finding no lenses.  The error bar (right, 
 offset) shows the maximum likelihood value and 90\% confidence range if the 
 PHFS sample were to contain one lens.  } 
\end{figure}

\begin{figure}
\centerline{\psfig{figure=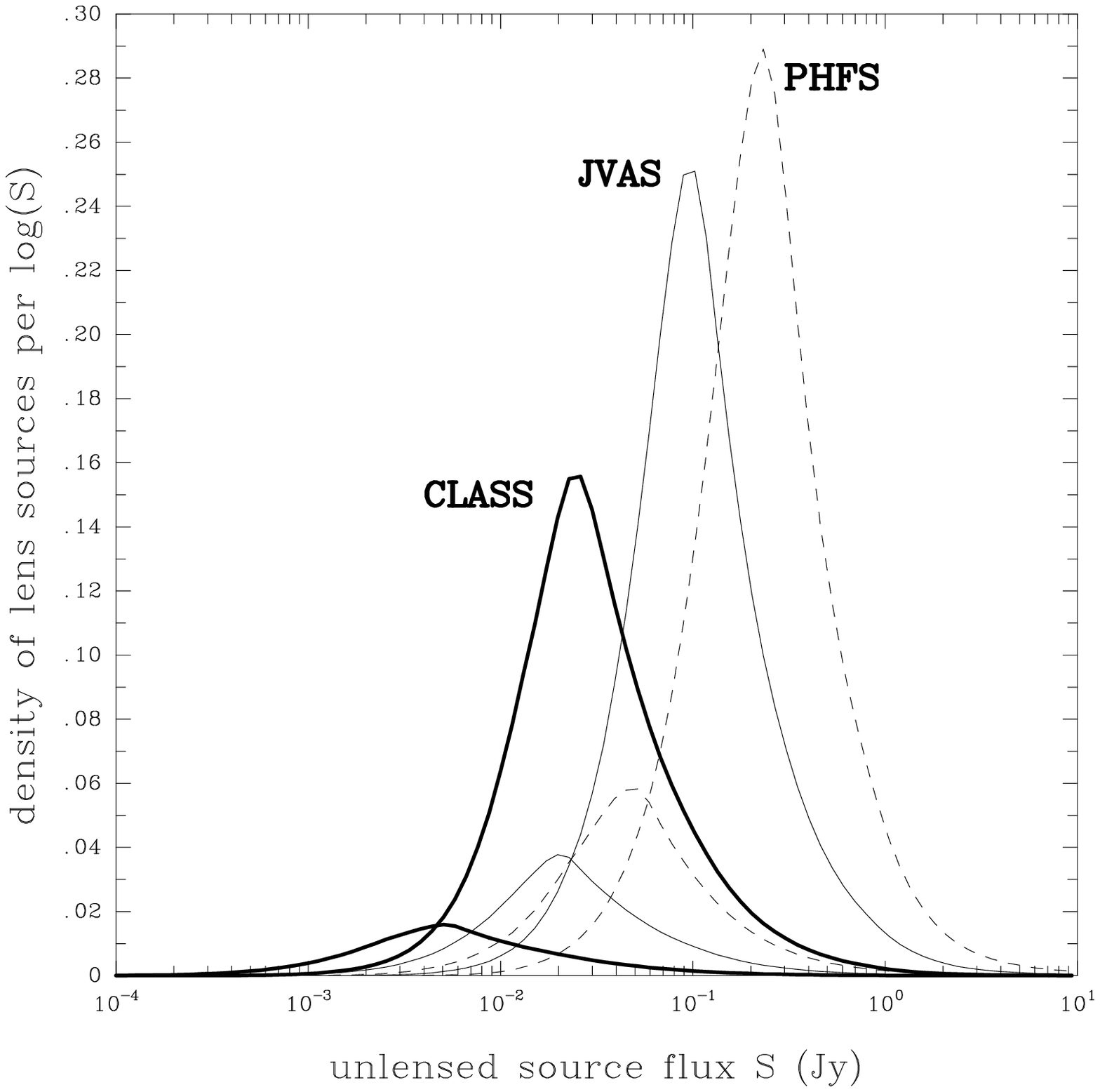,height=3.5in}
            \qquad
            \psfig{figure=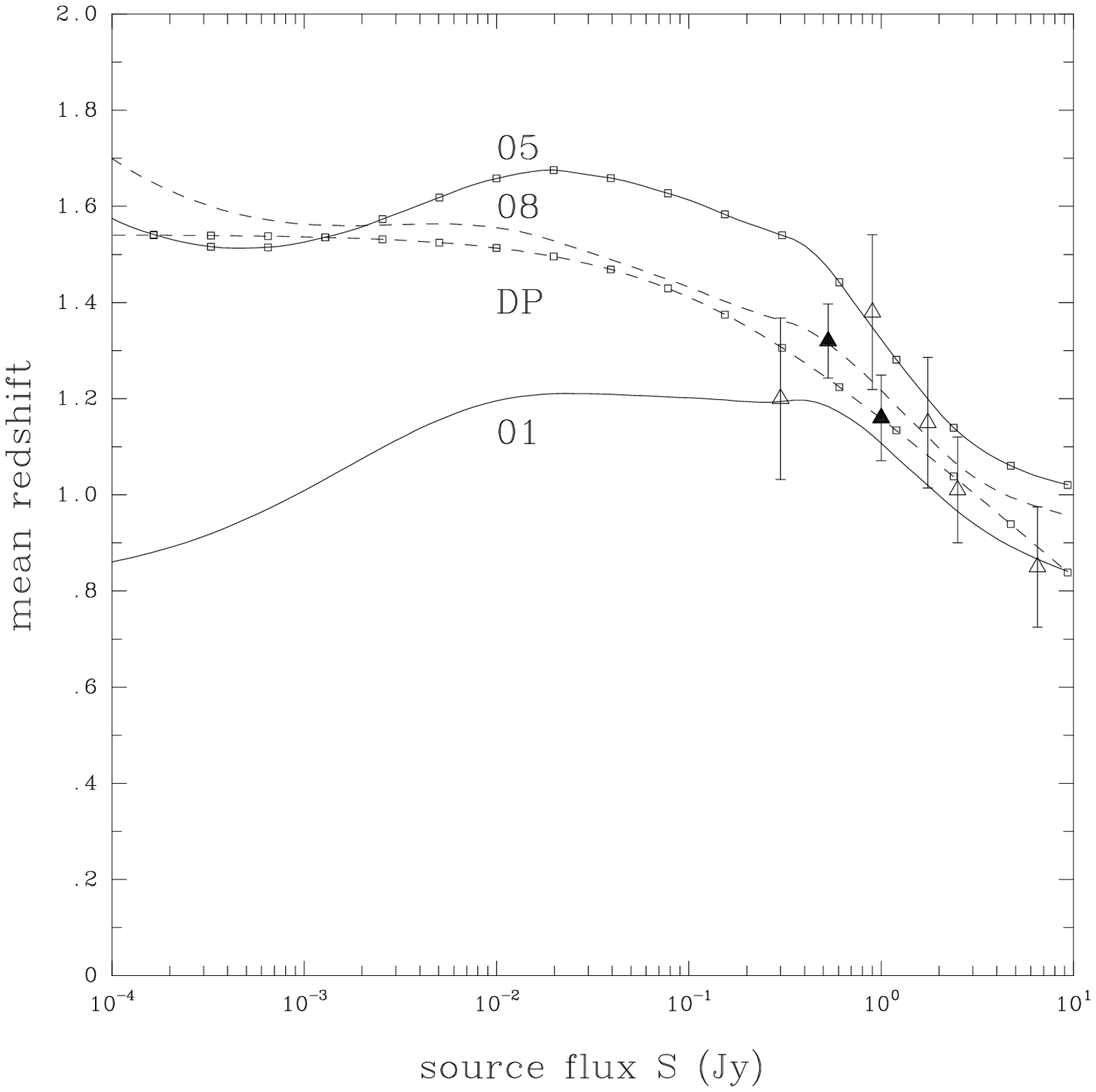,height=3.5in}}
\caption{ (Left) The unlensed flux distribution of lensed sources  for the PHFS
(dashed), JVAS (solid), and CLASS (heavy solid) samples in Model 01. The
higher peak in each pair is for the two-image systems, and the lower peak is
for the four-image systems.  The vertical scale is the number of lenses
per $10^3$ sources per $\log S$.}
\caption{ (Right) Mean source redshifts as a function of source flux for
Model 01 (solid), 05 (solid/points), 08 (dashed), and Dunlop \& Peacock (1990)
(dashed/points).  The error bars show the mean redshift and its uncertainty
in the existing redshift surveys, where the open triangles are the distributions
from Dunlop \& Peacock (1990), and the solid triangles are the CJI and CJII
distributions.  The curve for Model 05 is the mean for the distribution
without the completeness corrections -- adding the corrections shifts the
curve downwards and brings it into better agreement with the data. }
\end{figure}

\begin{figure}
\centerline{\psfig{figure=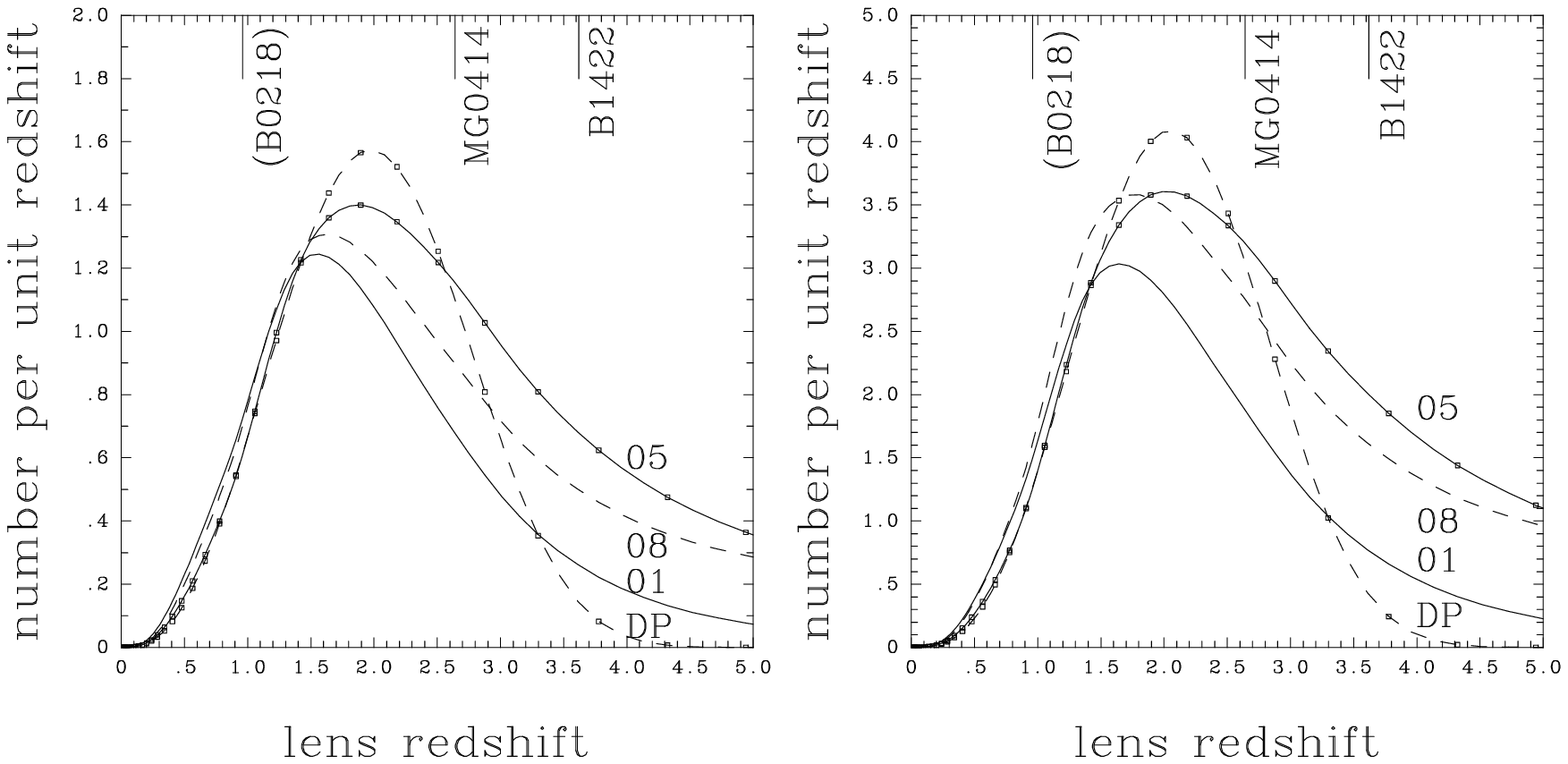,height=3.0in}}
\caption{ Lens (source) redshift distributions in the JVAS survey for
$\Omega_0=1$ (left) or $\Omega_0=0.35$ (right) flat cosmological models.
The distributions are shown for Model 01 (solid), 05
(solid/points), 08 (dashed), and Dunlop \& Peacock (1990) (dashed/points).
The known source redshifts are marked and labeled at the top (the redshift
of B0218 is tentative).}
\bigskip
\centerline{\psfig{figure=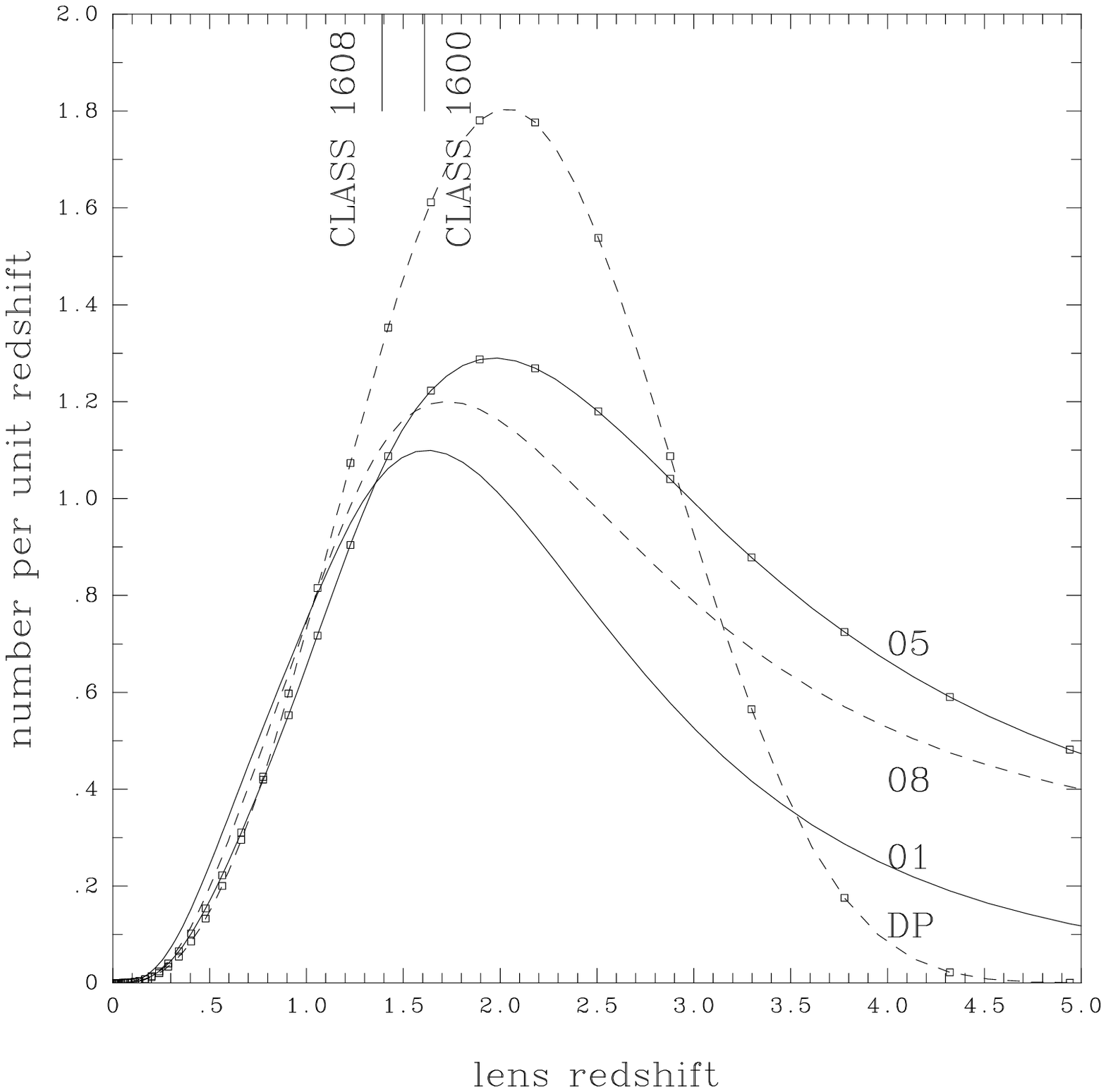,height=3.0in}}
\caption{ Lens (source) redshift distributions in the CLASS survey for
$\Omega_0=1$.  The distributions are shown for Model 01 (solid), 05
(solid/points), 08 (dashed), and Dunlop \& Peacock (1990) (dashed/points).
The known source redshifts are marked and labeled at the top.}
\end{figure}

We now want to explore methods of distinguishing the various RLF
models so that the cosmological uncertainties can be reduced.  The
direct approach is to complete old or conduct new redshift 
surveys of radio sources.  Figure 4 shows
the {\it unlensed} flux distribution of lensed sources in Model 01 
for the three surveys.
The distribution shows little variation with model.  In each 
case, the peak of the distribution lies near one-half of the flux limit 
for the two-image systems, and near one-tenth of the flux limit for the four-image 
systems.  The bulk of the lenses seen in the JVAS sample come from sources 
that are fainter than the existing redshift surveys ($S < 300$ mJy), and those 
with brighter sources lie in the region where the redshift surveys are 
incomplete ($S < 500$ mJy).  Almost all the CLASS lenses are 
from sources fainter than the flux limits of existing redshift surveys. 
Figure 5 shows the
variation of the mean source redshift with flux for several of the models, as
well as the constraints from redshift surveys of bright sources.  The numbers
of lenses found in the various JVAS models are strongly correlated with the
mean redshifts of 50-200 mJy sources.

The other way to reduce the uncertainties in the RLF is to use the
properties of the observed lenses,  although using the lenses adds 
many complications because their properties combine the effects of the RLF, 
the cosmological model, and the properties and distribution of the lens galaxies.  
Nonetheless, the flux and redshift distributions of the lenses may be as useful a
constraint on the RLF as direct surveys once there are larger numbers of
lenses.  

The flux distributions of the lenses in the various
models and different cosmologies are very similar and consistent with
the observed flux distribution of the lenses.  The similarity arises
because the lens probability curves (see Figure 3) have almost
identical shapes as a function of flux.  When normalized to
the observed number of lenses, the integral distribution of lens
fluxes are indistinguishable.  Even over longer flux baselines
(e.g. JVAS versus CLASS), the relative numbers of lenses show little 
change from model to model.

The redshift distributions, however, differ markedly, as 
Figures 6 and 7 show.  Models predicting larger numbers of lenses
generally have higher lens redshifts, independent of the cosmological
model.  Models where the completeness decreases with source redshift
(e.g. Model 05) allow many high redshift lenses ($z>4$).  All our
models predict some $4 < z < 5$ lenses, in contrast to the sharp
cutoff in the Dunlop \& Peacock (1990) model.  The smoothing 
function we use does not truncate the number density of high redshift
sources as abruptly simply because a more gradual reduction in 
the number of sources is both smoother and generally compatible with the
absence of such sources in the existing redshift surveys.  
For example, in Model 08 (05), the total
number of objects expected in the highest redshift bin ($4 < z < 5.16$)
for all seven surveys is 4.6 (3.0) and the surveys found no objects.  The next highest
redshift bin ($3.5 < z < 4$) has a predicted content
of 4.1 (3.3) objects and the surveys found 4 objects.   Although the
expected numbers of objects are similar, the source density in the higher
redshift bin is more than two times lower because of the change in the
bin widths.  For comparison, the Dunlop \& Peacock (1990) model predicts
$0.034$ sources in the high redshift bin and $0.45$ sources in the next
lower redshift bin.  The Poisson likelihoods of the source counts in these
bins for Model 08, Model 05, and Dunlop \& Peacock (1990) are 0.2\%, 0.9\%,
and 0.1\% respectively, compared to a peak Poisson likelihood of 20\%.
Our distributions decline too slowly, but the Dunlop \& Peacock (1990)
distribution declines too rapidly. Moreover, the whole problem may be
dominated by redshift dependent incompleteness.
The lensing optical depth rises steeply with redshift, so the lensed
sample will show a weaker drop off with source redshift than the 
unlensed sample.   While the $z>4$ sources represent 1\% or less
of the sources in the redshift surveys, they can be
10-20\% of the lensed sources if the cutoff in the RLF at high
redshift is less steep than the Dunlop \& Peacock (1990) model predicts.
 The three redshifts in the   
JVAS sample are compatible with almost any of these distributions,
except possibly the Dunlop \& Peacock (1990) distribution where
the $z=3.62$ redshift of B 1422+231 is unlikely.  Figure 7 shows
the expected lensed source redshift distribution for $\Omega_0=1$ in the 
CLASS survey.  The distributions are little changed from the JVAS
distribution.  The continued absence (or discovery) of high redshift 
lensed sources ($z > 4$), may be a more powerful constraint on the
high redshift RLF than direct redshift surveys.

\section{Models Constrained to Find Fixed Numbers of Lenses}

While we explored a range of 
smoothing terms in \S4, they may not be the terms to find the maximum, 
physically allowable range for the number of lenses.  For example, none of 
the nine models in Table 1 reached four observable lenses in the JVAS sample.  
We find little difference if we give the lenses their true statistical weight 
($\lambda_L=1$) because there are so few lenses.   We can, however, regard
the lensing terms as a type of smoothing by setting $\lambda_L=10^4$, 
and then converging the model until $L_{data}=147$ by adjusting the
Lagrangian multiplier of the smoothing $\lambda_U$.  In this section we
explore the properties of RLF models constrained to produce exactly two,
four or six observable lenses in the JVAS survey after correcting for 
resolution and including only compact sources for flat cosmological models 
with $0 < \Omega_0 < 2$.  

Although constrained to have $L_{data}=147$, the models do not necessarily 
agree with all the individual constraints.  If one data 
set is poorly fit because it cannot be satisfied given the lensing 
constraint, the target value of $L_{data}$ can still be reached by 
overfitting the other data.\footnote{We can avoid this by using separate 
Lagrange multipiers for each constraint, and then adjusting these multipliers 
until each individual constraint has a satisfactory fit.}  
For the models constrained to find four lenses, 
the poorly fit data are always the two lowest flux redshift distributions, 
the CJII sample and the Parkes Selected Area Sample.   For small matter
densities, the mean redshift is too low to agree with the data, and for high 
matter densities it is too high. Figure 8 shows the likelihood ratios for 
fitting the CJII and PSAS samples as a function of $\Omega_0$.  We can
find models with exactly four lenses that are consistent with these redshift
distributions for $0.5 \ltorder \Omega_0 \ltorder 1.4$ at 99\% confidence
if we simply use the likelihood ratio to determine the confidence intervals.
In short, finding four lenses in 2200 sources is compatible with
a substantial positive (negative) cosmological constant only if
the true lensing rate is significantly higher (lower) than in the
current sample.

\begin{figure}
\centerline{\psfig{figure=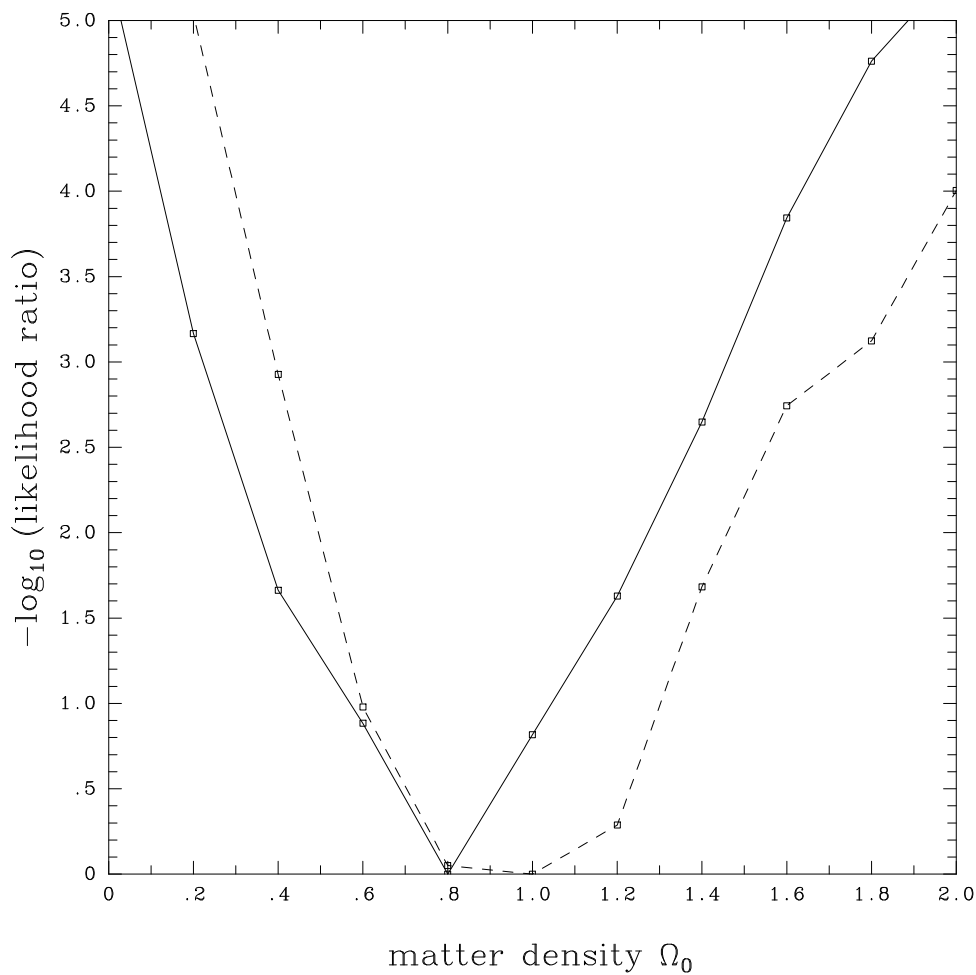,height=5.0in}}
\caption{The likelihood ratio $\log L/L_{max}$ of fitting the
 CJII and PSAS redshift distributions as a function of the matter
 density $\Omega_0$ in flat cosmologies for models constrained
 to have 4 JVAS lenses. The solid line uses the smoothing parameters
 of Model 01, and the dashed uses those of Model 05. } 
\end{figure}

Since the models are constrained to fit the observed number of
lenses exactly and are compatible with the flux and redshift constraints except 
for the most extreme cosmological models, we must find some other means of 
distinguishing between models.  When the number of JVAS lenses is fixed,
the numbers of PHFS and CLASS lenses are also nearly fixed (except for
$\Omega_0 \ltorder 0.2$).  The differences between models are found
in the redshift distributions of lensed and unlensed sources.

Both redshift distributions have strong variations with the cosmological
model when the number of lenses is fixed.
Figure 9 shows the redshift distribution of lensed sources
for models constrained to have four lenses (Model 01 smoothing terms)
as a function of the matter density.  Since the number counts data
constrain the total number of sources at any given flux, and the
shape of the magnification probability distribution is independent 
of the cosmological model, the only way to hold the number of 
lenses fixed is by adjusting the redshift distribution.
Models with a large positive cosmological constant have high optical 
depths, driving down the mean redshift, while models with a  large
negative cosmological constant have low optical depths, driving up
the mean redshift.  The same effect can be seen in the mean
redshift of unlensed sources shown in Figure 10.    
We can also search for models in a fixed cosmology producing different
numbers of lenses.  Figures 9 and 10 also show the properties of
$\Omega_0=1$ models (with Model 01 smoothing terms) that produce either
two or six lenses in the JVAS sample.  The two lens model is consistent
with all the RLF constraints, but the six lens model fits the
PSAS redshift data poorly.   

\begin{figure}
\centerline{\psfig{figure=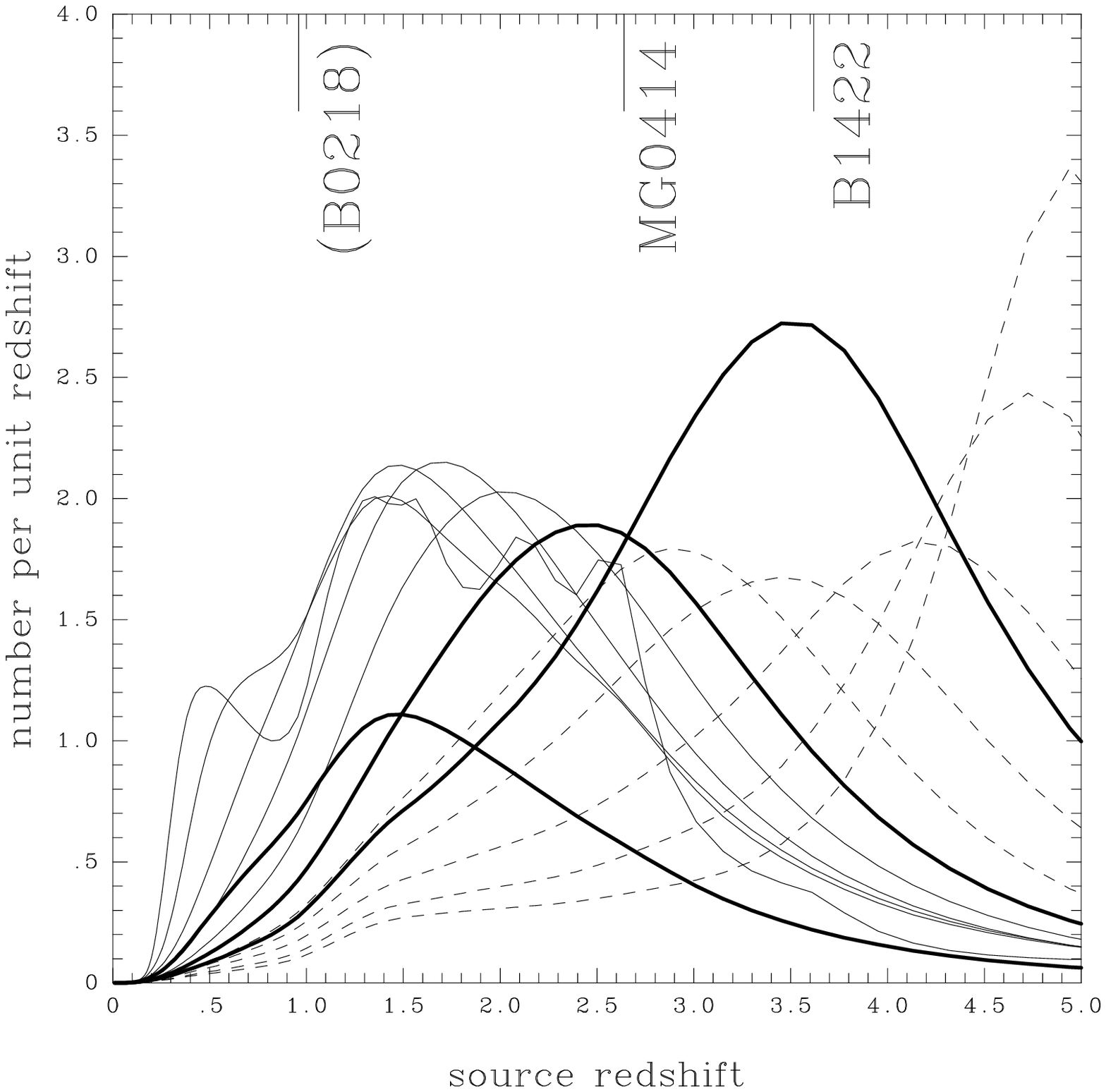,height=3.5in}
            \qquad
            \psfig{figure=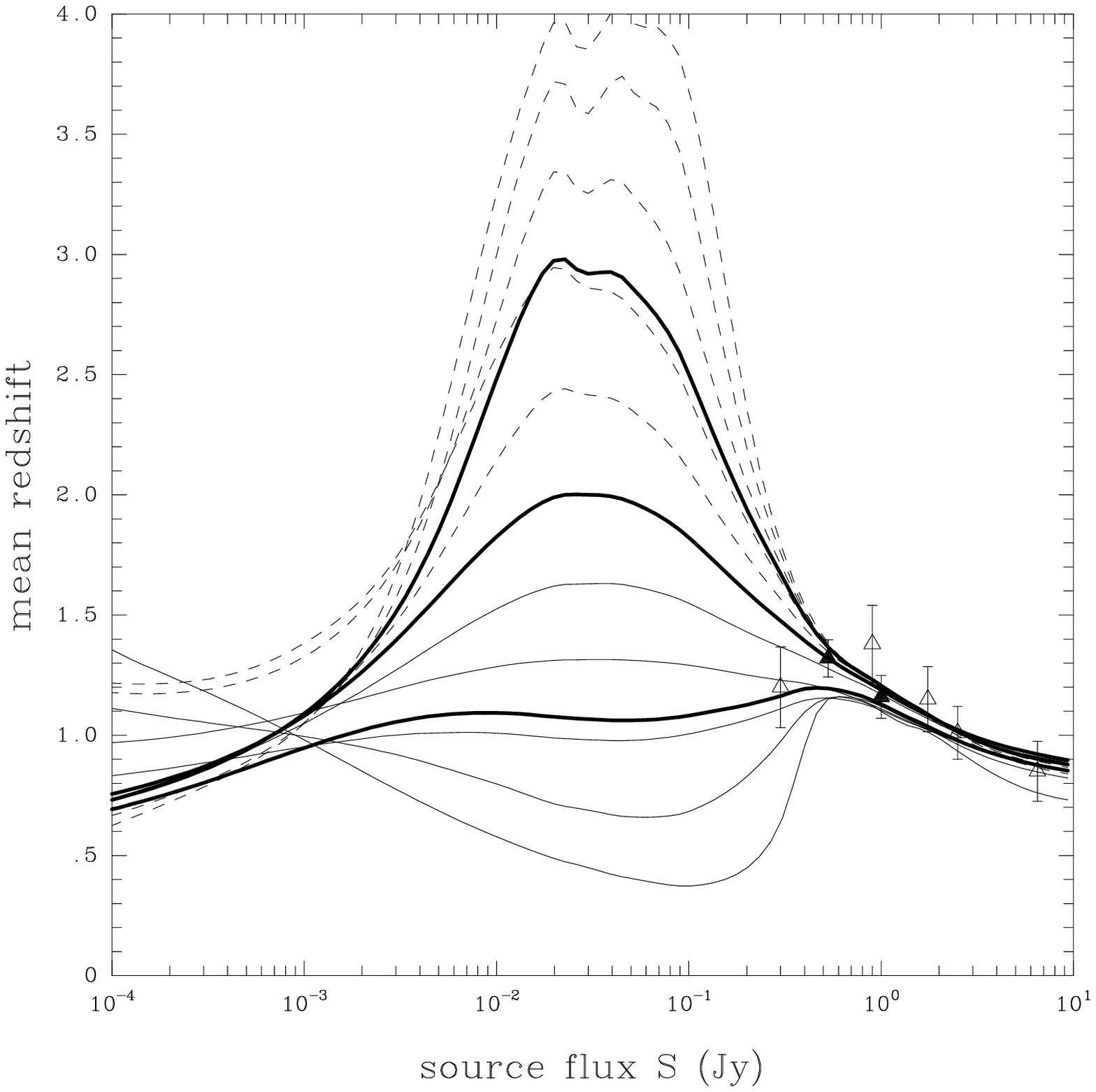,height=3.5in}}
\caption{(Left) The redshift distribution of lensed sources constrained
 to have a fixed number of JVAS lenses. The heavy solid lines have $\Omega_0=1$
 and produce 2, 4, or 6 lenses in order of increasing mean redshift.
 The light solid lines with peaks shifting to lower redshifts are constrained
 to produce four lenses and have $\Omega_0=0.8$, $0.6$, $0.4$, 
 $0.2$, and $0.0$, while the dashed lines with peaks shifting to
 higher redshifts are constrained to produce four lenses and 
 have $\Omega_0=1.2$, $1.4$, $1.6$, $1.8$, and $2.0$,
 all with $\Omega_0+\lambda_0=1$.  The wiggles in the low $\Omega_0$
 models are caused by the overfitting needed to reach
 $L_{data}=147$.  We used the Model 01 smoothing terms. 
 }
\caption{(Right) The mean redshift of unlensed sources as a function
of source flux for RLF models constrained to produce a fixed number
of JVAS lenses using the Model 01 smoothing terms.  The mean redshifts 
from the seven redshift constraints are shown by the points.  
 The heavy solid lines have $\Omega_0=1$ and 2, 4, or 6 lenses
 in order of increasing mean redshift, the solid lines with peaks
 shifting to lower redshifts have 4 lenses and $\Omega_0=0.8$, $0.6$, $0.4$, 
 $0.2$, and $0.0$, while the dashed lines with peaks shifting to
 higher redshifts have 4 lenses and $\Omega_0=1.2$, $1.4$, $1.6$, $1.8$, and $2.0$,
 all with $\Omega_0+\lambda_0=1$.  
 }
\end{figure}

This experiment, using the lensing constraints as a type of smoothing term, demonstrates
that the range for the expected number of lenses in the JVAS survey is broader than
the range we found for the models in Table 1.  There are plausible, consistent RLF
models that produce 4 observable lenses in the JVAS sample for $\Omega_0=1$.  Most
of the uncertainty can be eliminated by determining the source redshift distribution
for fluxes between 10 and 300 mJy.  Completing the surveys in the 300 to 1000 mJy 
flux range (the CJI, CJII, and PSAS surveys) will help, but most lenses in the JVAS
sample come from the fainter flux range.

\section{The ``Ellipticity Crisis''}

King \& Browne (1996) point out that the JVAS sample contains an anomalously
high number of four-image lenses, and that there are signs of a similar
problem in the CLASS sample.  The observed fraction (again neglecting
B 1938+666, which has both two- and four-image parts) is one in two, while the
predicted ratio is one in five for our standard model.  The two issues we 
must consider are whether the problem exists, and how to solve it.

\begin{figure}
\centerline{\psfig{figure=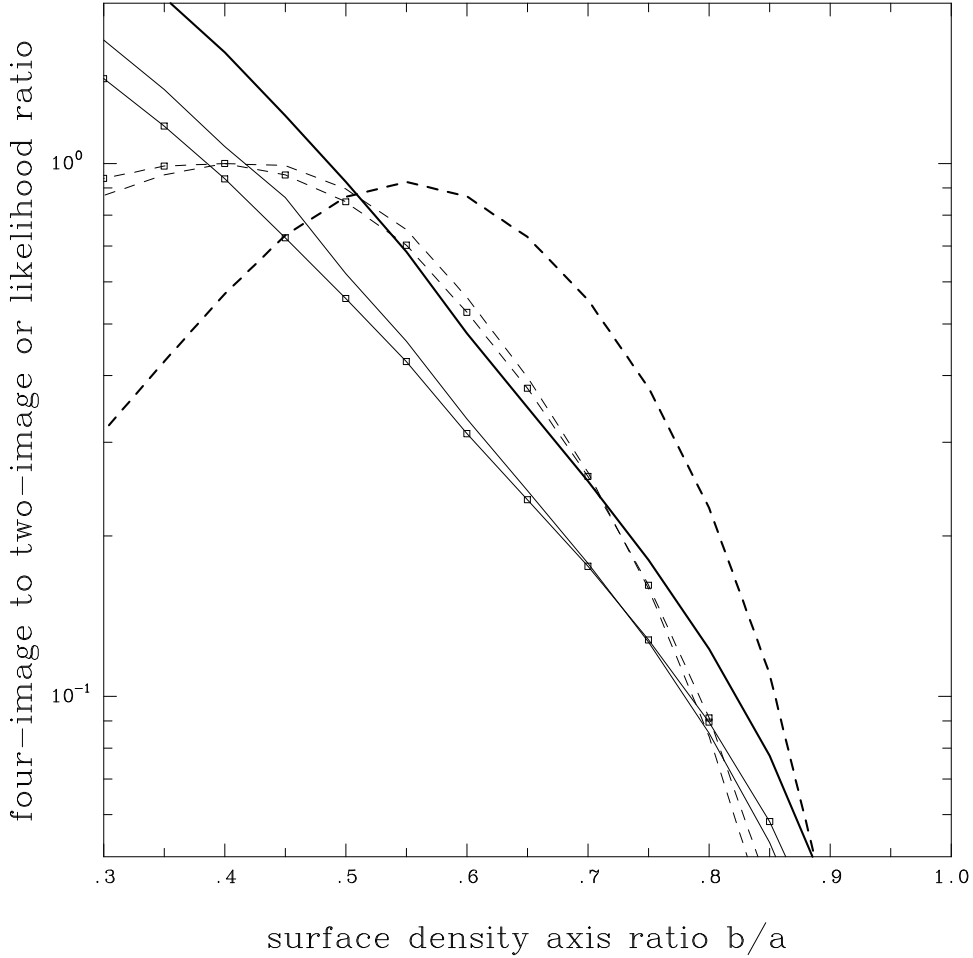,height=5.0in}}
\caption{The four-image to two-image number ratio as a function of galaxy axis 
ratio $b/a$ (solid curves) and the corresponding maximum likelihood ratios 
(dashed curves).  The light solid curves show the ratio for Model 01 (no points)
and a model with 4 lenses and the Model 01 smoothing curves (with points), 
and the heavy solid line shows the equivalent ratio for the quasar surveys 
discussed by Kochanek (1996).  The quasar surveys contain 3 two-image and 2 
four-image 
lenses, while the JVAS survey contains 2 two-image and 2 four-image lenses of 
compact sources.  The corresponding dashed lines show the likelihood ratio for 
the models to produce the observed numbers of two- and four-image systems when 
constrained to have the observed total number of lenses.  
 }
\end{figure}

Does a problem exist? For Model 01 in \S4, including the separation completeness 
factor and dropping the extended sources, we expect $1.92$ two-image and $0.34$ 
four-image lenses in the observed sample.  
The Poisson probability of finding at least as many four-image as two-image
lenses given these expectation values is 24\% if we do not constrain the
distributions to have four lenses, and 11\% for samples containing four lenses.
For Model 05, we expect $2.83$ two-image and $0.52$ four-image lenses,
and the probability of finding at least as many four-image as two-image
systems is 15\% if unconstrained, and 12\% if constrained to have four images 
(see Figure 3).  In the $\Omega_0=1$ model constrained to have four lenses (see \S5)
we expect 3.36 two-image and 0.62 four-image systems, with a 12\% chance of having at
least as many four-image as two-image systems.

So far, all models were computed using the statistical predictions from
a uniform distribution of galaxies with axis ratios between $b/a=0$ and 
$b/a=0.5$.  This ``standard'' model has a two-image to four-image number ratio 
comparable to the ratio for a model with an axis ratio between $b/a=0.65$ and
$0.70$ (see Figure 1). Figure 11 shows the 
expected ratio of four-image to two-image lenses in the JVAS survey as a 
function of the axis ratio of the surface density, as well as the equivalent 
results for the quasar lens sample used by Kochanek (1996), which contains 
three two-image and two four-image lenses.  The quasars have more four-image 
lenses for a given ellipticity because the magnification bias of the
sample is greater (see Kochanek 1991, Wallington \& Narayan 1993, Kassiola \& 
Kovner 1993).  To match the observed JVAS ratio of 1:1 requires $b/a=0.4$, much
more flattened than a typical elliptical galaxy (see Ryden 1992).  We quantify 
the agreement using the maximum likelihood ratio to estimate confidence limits 
from the Poisson probability of finding 2 two-image and 2 four-image lenses 
in a sample containing four lenses in total.  The largest axis ratios agreeing 
with the data at the 1--$\sigma$, 90\% confidence, and 2--$\sigma$ confidence 
limits have $b/a\simeq 0.55$, $b/a\simeq 0.70$, and $b/a\simeq 0.75$ 
respectively.  Our standard model for the ellipticity distribution lies near 
the 90\% confidence limit.  

The quasar model matches the observed ratio at an axis ratio of $b/a=0.55$, 
which is not as elliptical as the best fit to the JVAS data, but more 
elliptical than our standard model.  The 1--$\sigma$ and 90\% maximum 
likelihood confidence limits for this model are 
$0.35 \ltorder b/a \ltorder 0.70$ and $0.25 \ltorder b/a \ltorder 0.75$.
The standard model is still too circular, but only at the 1--$\sigma$ level. 
The radio and quasar ellipticity probability distributions are perfectly 
compatible, and if we examine the joint probability distribution we find that 
the best fit is $b/a=0.50$, with a 1-$\sigma$ range of 
$0.35 \ltorder b/a \ltorder 0.60$ and a  90\% confidence range of 
$0.25 \ltorder b/a \ltorder 0.75$.  The standard model 
lies between the 1-$\sigma$ and 90\% confidence limits.

In short, the observed ratio of two-image to four-image lenses is somewhat, 
but not significantly,
unlikely.  There are, however, other signs of ellipticity problems from 
highly elliptical lens models or gross disagreements between the models and
the observed lens galaxy  (see Kochanek (1991), or Ratnatunga et al. 1995
for some examples), a point strongly emphasized by Schechter (1996).
Some solutions to the statistical ellipticity problem help the model ellipticity
problem, but not all.

The statistical discrepancy can be alleviated by adding a core radius to the
lens models, which will mostly affect the numbers of two-image systems, or by
unmodeled systematic biases in the surveys against recognizing two-image 
systems.  These two effects reduce the number of two-image lenses while 
leaving the number of four-image lenses unchanged for a fixed RLF model.
However, the systematic uncertainties associated with the RLF can easily 
compensate for the variations 
in the expected number of lenses to keep the models in agreement with 
plausible cosmological models (as demonstrated in \S5).  
These are viable solutions to the statistical 
ellipticity problem, but they cannot 
explain high ellipticities in individual lens models.

We can also use the RLF to increase the number of four-image systems.
Figure 4 shows that the two- and four-image systems in a particular survey
are produced by sources with different fluxes.  By boosting the 
number of faint sources in the four-image peak relative to the number of 
sources in the two-image peak, we increase the relative number of four-image 
systems.  This effect causes the higher ratio of four- to two-image systems 
predicted in the quasar surveys (see Figure 11), because the
bright quasar ($m \ltorder 18$ B mags) LF is steeper than the RLF.
Note, however, that the four-image source peak for the JVAS survey lies in the 
middle of the two-image source peak for the CLASS survey.  Models boosting the 
ratio in the JVAS survey usually reduce the ratio in the CLASS survey.  
If we change the likelihood for the lensing model to separately fit the numbers
of two- and four-image systems, we can significantly increase the ratio.

We have not included spiral galaxies in our calculations, but they cannot solve
the ellipticity problem even if they are very flattened.  Current evidence
(see Sackett 1996) suggests that spiral halos are nearly axisymmetric and 
oblate, with a three-dimensional axis ratio of $\sim 0.5$.  The projected 
two-dimensional axis ratios will be higher.  Spiral galaxies produce only a
fraction $x=0.15$ to $0.20$ of the optical depth produced by the early type 
galaxies -- their lower average mass outweighs their higher number density 
(see Fukugita \& Turner 1990, Maoz \& Rix 1993, Kochanek 1993, 1996).  If 
fractions $r_e$ and $r_s$ of E/S0 and spiral lenses are four-image lenses 
respectively, then the overall fraction of four-image lenses is
$r = (r_e + x r_s)/(1+x) \simeq 0.83 r_e + 0.17 r_s$ (for $x=0.2$).  If we use 
the canonical $r_e=0.16$ from our typical RLF models, then 
$r_s = 0.5 + 0.35/x = 2.25$ for $x=0.2$ is needed for the spirals to explain 
equal numbers of two- and four-image lenses.  Such a value for $r_s$ implies 
a surface density axis ratio smaller than $b/a=0.3$, contradicting the
dynamical estimates of the properties of spiral halos.  Thus, spiral galaxies 
are of little help in explaining the ellipticity problem.


\section{Conclusions}

The results of the JVAS lens survey (see Patnaik 1994, King \& Browne 1996) are 
consistent with models for the 
statistics of lensed quasars (Kochanek 1996), although the cosmological uncertainties are 
broader because of the systematic uncertainties in the RLF (radio luminosity function). 
This agreement is a remarkable affirmation of lens statistical models, since the only
common assumption is the model for the number and mass distribution of lens galaxies.
There are four JVAS lenses produced by compact sources, and in an $\Omega_0=1$ cosmological 
we predicted between 2.3 and 3.4 lenses for a series of RLF models consistent with 
observational constraints.  In fact, there is no problem finding a model RLF that produces exactly four
observable lenses in this sample for $\Omega_0=1$.  In flat cosmological
models ($\Omega_0+\lambda_0=1$) the systematic uncertainties are consistent with a
broad range of values for the matter density, with $0.20 \ltorder \Omega_0 \ltorder 2.0$
at 90\% confidence. 

In addition to the JVAS survey, we estimated the number of lenses expected in the
CLASS and PHFS surveys.  So far, the CLASS survey (Myers 1996, Myers et al. 1995,
Jackson et al. 1995) has found two lenses in a larger,
fainter sample than the JVAS survey.  Our models predict that the number of lenses
expected in the first part of the CLASS survey is approximately equal to the number 
expected in the
JVAS survey.  Although the CLASS survey contains more sources, the lensing probability
is a declining function of source flux.  The PHFS survey (Webster et al. 1996) is examining 
a much smaller sample of brighter sources, and our models predict only 0.3 to 0.6 lenses for
$\Omega_0=1$.  There are too few lensed source redshifts to distinguish between RLF 
models in the current samples, but they are a promising means of constraining models.  
The flux distribution of the lensed sources is consistent with the model predictions.

The systematic uncertainties in the expected number of lenses and the RLF model
are created by the absence of redshift information for (unlensed) sources fainter
than 300 mJy.  The source of a typical JVAS lens is a 50--200 mJy source, and the
source of a typical CLASS lens is a 10--50 mJy source, well below the fluxes
with even partial redshift distributions.  The redshift distribution of the sources
at these fluxes is largely
constructed from assumptions about smoothness and evolution, and the resulting
uncertainty in the mean redshift of the sources permits large variations in the
expected number of lenses.  If, for example, we examine model RLFs that produce
exactly 4 JVAS lenses, the mean redshift of a 50 mJy source varies from 0.4
for $\Omega_0=0$, to 1.9 for $\Omega_0=1$, to almost 4.0 for
$\Omega_0=2$ in flat cosmologies with a cosmological constant ($\Omega_0+\lambda_0=1$).
The extremes for the mean redshift are implausible, but they serve to illustrate
the direct relation between mean source redshift and cosmological uncertainties.
{\it Using the flat spectrum radio lens surveys
to study the cosmological model depends on determining the redshift distribution
of radio sources in the flux range from 10 to 300 mJy.}  
While it is helpful to complete the existing brighter redshift surveys (CJI and CJII
(e.g. Taylor et al. (1994), Henstock et al. (1995), Thakkar et al. (1995)),
PHFS (Parkes Half-Jansky Flat-Spectrum Survey, Webster et al. 1996), and PSAS (Parkes
Selected Areas Survey, Dunlop et al. 1986, Dunlop et al. 1986, Allington-Smith et al. 1991) 
samples), they are at the wrong fluxes to eliminate the cosmological uncertainties.
Because the redshift variations are so large, relatively small (50 sources) redshift
surveys of modest completeness (80\% or better) can eliminate most of the uncertainties.

As noted by King \& Bowne (1996), the number of four-image lenses in the JVAS survey
(2 of 4 compact-source lenses) is significantly greater than the 14-16\% predicted
by theoretical models using ellipticities typical of E and S0 galaxies.  A uniform 
distribution of lenses in axis ratio from $b/a=0.5$ to $b/a=1.0$ (roughly the
distribution for E and S0 galaxies, e.g. Schechter (1987), Ryden (1992)) produces the same
numbers of four-image lenses as a galaxy with an axis ratio of $b/a\simeq 0.65$.  A model
that produces the observed ratio of four-image to two-image lenses is too elliptical,
with an axis ratio of $b/a\simeq 0.4$.
The discrepancy is significant only at the 90\% confidence level, since the 1-$\sigma$,
90\% confidence, and 2-$\sigma$ upper limits on the axis ratio are $b/a=0.55$,
$0.70$, and $0.75$ respectively.  Fitting the number of four-image lenses 
found in quasar lens surveys (see Kochanek 1991, 1996, Wallington \& 
Narayan 1993, Kassiola \& Kovner 1993) also requires higher than expected 
ellipticities, but the
discrepancy is significant only at the 1-$\sigma$ confidence level. The 
ellipticity estimates for the radio and quasar lenses are perfectly compatible,
and the joint distribution has a best fit axis ratio of $b/a=0.5$ and a
90\% confidence range of $0.25 \ltorder b/a \ltorder 0.65$.   While the
best fitting lens models are more elliptical than expected for E/S0 galaxies,
the Poisson uncertainties are so broad that the significance of the discrepancy
is only at the 90\% confidence level. 

There are four plausible solutions to the ellipticity
problem.  The first solution is that the problem does not exist, since the statistical
significance of the disagreement is low.  The second solution is that systematic
errors in the calculations lead to the discrepancy.  For example, there are many
reasons our selection effects model may overestimate the detectability of two-image 
lenses.  The RLF allows enough freedom in the total number of lenses to halve the 
number of detectable two-image lenses and still remain consistent with the lensing and
RLF constraints. 
Schechter (1996) emphasizes that there is also a problem with models of individual
lenses being too elliptical, and neither of these solutions to the statistical 
ellipticity problem can address this.  The third solution is that the dark
matter halos of early type galaxies are more elliptical than the light, and
the fourth solution is that external shear perturbations from structure near
the lens or along the line of sight augment the intrinsic ellipticities of the
lens galaxies (e.g. Kochanek \& Apostolakis 1988, Bar-Kana 1996). 

\acknowledgements  Acknowledgements: The author thanks P. Schechter for 
discussions and E. Falco for comments.  The research was supported by the
Alfred P. Sloan Foundation and NSF grant AST 94-01722.


\begin{references}

  \def\aa{{ A\&A}}
  \def\aj{{ AJ}}
  \def\annrev{{ ARA\&A}}
  \def\apj{{ ApJ}}
  \def\apjl{{ ApJ}}
  \def\apjs{{ ApJS}}
  \def\mn{{ MNRAS}}
  \def\Nature{{ Nature}}
  \def\science{{ Science}}
  \def\pasp{{ PASP}}
  \def\rmp{{ RevModPhys}}
\def\refindent{\par\penalty-100\noindent\parskip=4pt plus1pt
               \hangindent=3pc\hangafter=1\null}
\def\ref#1#2#3#4{\refindent#2, {#1\/,\ }{#3}, #4}
\def\reft#1#2#3#4#5#6{\refindent#2 {\it #6}, #3, {\it #1\/,\ }{\bf#4}, #5.}
\def\reftb#1#2#3#4#5#6{\refindent#2 #3, {\it #6}, {\it #1\/,\ }{\bf#4}, #5.}
\def\refbook#1{\refindent#1}
\def\preprint#1#2#3{\refindent#1, #2, {\it #3 preprint}}
\def\preprintt#1#2#3#4{\refindent#1, #2 {\it #4}, {\it #3 preprint}.}
\def\refinpress#1#2{\refindent#1, {\it #2, in press}.}
\def\reftinpress#1#2#3#4{\refindent#1 {#4}, #2, {\it #3, in press}}
\def\refsubmit#1#2{\refindent#1, {\it #2, submitted}}
\def\reftsubmit#1#2#3#4{\refindent#1 {#4}, #2, {\it #3, submitted}}
\def\reftinprep#1#2#3{\refindent#1 {#3}, #2, {\it in preparation}}

\ref\mn{Allington-Smith, J.R., Peacock, J.A., \& Dunlop, J.S., 1991}{253}{287}
\ref{A\&ASuppl}{Altschuler, D.R., 1986}{65}{267}
\preprint{Bar-Kana, R.}{1996}{astro-ph/9511056}
\ref\apj{Bennett, C.L., Lawrence, C.R., \& Burke, B.F., 1985}{299}{373}
\ref\apj{Blandford, R.D., \& Narayan, R. 1986}{310}{568}
\ref\aa{Breimer, T.G., \& Sanders, R.H., 1993}{274}{96}
\refbook{Burke, B.F., Leh\`ar, J., \& Conner, S.R., 1992, in Gravitational 
  Lenses, ed. R. Kayser,  T. Schramm, \& L. Nieser  (Springer: Berlin) 237}
\ref\annrev{Carroll, S.M., Press, W.H., \& Turner, E.L., 1992}{30}{499}
\ref\apj{Condon, J.J., 1989}{338}{13}
\ref\aj{Condon, J.J., \& Ledden, J.E., 1981}{86}{643}
\ref\apj{Donnelly, R.H., Partridge, R.B., \& Windhorst, R.A., 1987}{321}{94}
\refbook{Dunlop, J.S., 1994, in Frontiers of Space and Ground-Based Astronomy,
  eds., W. Wamsteker et al. (Kluwer: Dordrecht) 395}
\ref\mn{Dunlop, J.S., Peacock, J.A., Savage, A., \& Carrie, D.R., 1986}{218}{31}
\ref\mn{Dunlop, J.S., Peacock, J.A., Savage, A., Lilly, S.J., Heasley, J.N., 
  \& Simon, A.J.B., 1989}{238}{1171}
\ref\mn{Dunlop, J.S., \& Peacock, J.A., 1990}{247}{19}
\preprint{Driver, S.P., Windhorst, R.A., Phillipps, S., \& Bristow, D.}
   {1995}{astro-ph/9511141}
\ref\apj{Faber, S.M., \& Jackson, R.E., 1976}{204}{668}
\ref{Science}{Fomalont, E.B., Kellerman, K.I., Wall, J.V., \& Weistrop, D., 1984}{225}{23}
\ref\aj{Fomalont, E.B., Windhorst, R.A., Kristian, J.A., \& Kellerman, K.I., 1991}{102}{1258}
\refbook{Franx, M., 1993, in  Galactic Bulges, ed. H. Dejonghe \& H.J. Habing
  (Dordrecht: Kluwer) 243}
\ref\mn{Fukugita, M., Futamase, T., \& Kasai, M., 1990}{246}{24p}
\ref\mn{Fukugita, M., \& Turner, E.L., 1991}{253}{99}
\ref\apjs{Gregory, P.C., \& Condon, J.J., 1991}{75}{1011}
\ref\apjs{Henstock, D.R., Browne, I.W.A., Wilkinson, P.N., Taylor, G.B., Vermeulen, R.C.,
  Pearson, T.J., \& Readhead, R.C.S., 1995}{100}{1}
\ref{Spectrum}{Henstock, D.R., Browne, I.W.A., \& Wilkinson, P.N., 1994}{4}{??}
\ref\Nature{Hewitt, J.N., Turner, E.L., Schneider, D.P., Burke, B.F.,
  Langston, G.I., \& Lawrence, C.R., 1988}{333}{537}
\ref\mn{Hogg, D.W., \& Blandford, R.D., 1994}{268}{889}
\ref\mn{Jackson, N., de Bruyn, A.G., Myers, S., Bremer, M.N., Miley, G.K., 
   Schilizzi, R.T., Browne, I.W.A., Nair, S., Wilkinson, P.N., Blandford, 
   R.D., Pearson, T.J., \& Readhead, A.C.S., 1995}{274}{L25}
\ref\apj{Kassiola, A., \& Kovner, I., 1993}{417}{450}
\refbook{Kellerman, K.I., \& Wall, J.V., 1987, in Observational Cosmology, eds., A. Hewitt,
  et al., (Kluwer: Dordrecht) 543}
\ref\mn{King, L.J. \& Browne, I.W.A., 1996}{282}{67}
\refbook{King, L.J., Browne, I.W.A.,  Wilkinson, P.N., \& Patnaik, A.R., 1996, 
  in Astrophysical Applications of Gravitational Lensing, eds.,
  C.S. Kochanek \& J.N. Hewitt (Kluwer: Dordrecht)  191}
\ref\apj{Kochanek, C.S., 1991}{379}{517}
\ref\aj{Kochanek, C.S., \& Lawrence, C.R., 1990}{99}{1700}
\ref\apj{Kochanek, C.S., 1993}{419}{12}
\ref\apj{Kochanek, C.S., 1994}{436}{56}
\refinpress{Kochanek, C.S., 1996}{ApJ}
\preprint{Lawrence, C.R., Elston, R., Jannuzi, B.T., \& Turner, E.L.}{1994}{Caltech}
\refbook{Leh\'ar, J., 1991, MIT PhD Thesis}
\ref\apj{Lilly, S.J., Tresse, L., Hammer, F., Crampton, D., \& Le F\`evre, O., 1995}{455}{108}
\ref\apj{Loveday, J., Peterson, B.A., Efstathiou, G., \& Maddox, S.J., 1992}{390}{338}
\ref\apj{Mao, S., 1991}{380}{9}
\ref\mn{Mao, S., \& Kochanek, C.S., 1994}{268}{569}
\ref\apj{Maoz, D., \& Rix, H.-W., 1993}{416}{425}
\ref\aj{Marzke, R.O., Geller, M.J., Huchra, J.P., \& Corwin, H.G., 1994}{108}{437}
\ref\aa{Maslowski, J., Pauliny-Toth, I.I.K., Witzel, A., \& K\"uhr, H., 1981}{95}{285}
\ref\apjl{Myers, S.T., Fassnacht, C.D., Djorgovski, S.G., et al., 1995}{447}{L5}
\refbook{Myers, S.T., 1996, in Astrophysical Applications of Gravitational Lensing, eds.,
  C.S. Kochanek \& J.N. Hewitt (Kluwer: Dordrecht)  317}
\ref\aj{Owen, F.N., Condon, J.J., \& Ledden, J.E., 1983}{88}{1}
\refbook{Patnaik, A.R., 1994, in  Gravitational
  Lenses in the Universe, ed., J. Surdej, D. Fraipont-Caro, E. Gosset, S. Refsdal,
  \& M. Remy (Li\`ege, Universit\'e de Li\`ege) 311}
\refbook{Patnaik, A.R., Browne, I.W.A., King, L.J., Muxlow, T.W.B., Walsh, 
  D., \& Wilkinson, P.N., 1992a, in {Gravitational Lenses}, ed. R. Kayser,
  T. Schramm, \& L. Nieser (Springer: Berlin) 140}
\ref\mn{Patnaik, A.R., Browne, I.W.A., Wilkinson, P.N., \& Wrobel, J.A., 1992}{254}{655}
\ref\aj{Pauliny-Toth, I.I.K., Witzel, A., Preuss, E., K\"uhr, H., Kellerman, K.I.,
  \& Fomalont, E.B., 1978}{83}{451}
\ref\mn{Peacock, J.A., 1983}{202}{615}
\ref\mn{Peacock, J.A., 1985}{217}{601}
\ref\mn{Peacock, J.A., \& Wall, J.V., 1981}{194}{331}
\ref\apjs{Polatidis, A.G., Wilkinson, P.N., Xu, W., Readhead, A.C.S., Pearson, T.J.,
  Taylor, G.B., \& Vermeulen, R.C., 1994}{98}{1}
\refbook{Press, W.H., Teukolsky, S.A., Vetterling, W.T., \& Flannery, B.R., 1992,
  Numerical Recipes (Cambridge Univ. Press: Cambridge) }
\ref\apjl{Ratnatunga, K.U., Ostrander, E.J., Griffiths, R.E., \& Im, M., 1995}{453}{L5}
\ref\apj{Rix, H.-W., Maoz, D., Turner, E.L., \& Fukugita, M., 1994}{435}{49}
\ref\apj{Ryden, B.S., 1992}{396}{445}
\refbook{Sackett, P.D., 1996, in Astrophysical Applications of Gravitational Lensing, eds.,
  C.S. Kochanek \& J.N. Hewitt (Kluwer: Dordrecht)  165}
\ref\apj{Schechter, P., 1976}{203}{297}
\refbook{Schechter, P., 1987, in Structure and Dynamics of Elliptical Galaxies, 
  ed. T. de Zeeuw (Kluwer: Dordrecht) 217}
\refbook{Schechter, P., 1996, private communication}
\ref\apjs{Taylor, G.B., Vermeulen, R.C., Pearson, T.J., Readhead, A.C.S., 
  Henstock, D.R., Browne, I.W.A., \& Wilkinson, P.N., 1994}{93}{345}
\ref\apjs{Thakkar, D.D., Xu, W., Readhead, A.C.S., Pearson, T.J., Henstock, D.R.,
  Browne, I.W.A., \& Wilkinson, P.N., 1994}{98}{33}
\ref\aa{Toffolatti, L., Franceschini, A., De Zotti, G., \& Danese, L., 1987}{184}{7}
\preprint{Tomita, K.}{1995}{YITP/U94-2}
\ref\apjl{Turner, E.L., 1990}{365}{L43}
\ref\mn{Wall, J.V., \& Peacock, J.A., 1985}{216}{173}
\ref\apj{Wallington, S., \& Narayan, R., 1993}{403}{517}
\refbook{Webster, R.L., Francis, P.J., Holman, B.A., Masci, F.J., Drinkwater, M.J.,
  \& Peterson, B.A., 1996, in Astrophysical Applications of Gravitational Lensing, eds.,
  C.S. Kochanek \& J.N. Hewitt (Kluwer: Dordrecht)  393}
\refbook{Windhorst, D., Mathis, D., \& Neuschaefer, L., 1990, in Evolution of the
  Universe of Galaxies, ed. R.G. Kron (Astron. Soc. of the Pacific: San Francisco) 389}
\ref\aj{Witzel, A., Schmidt, J., Pauliny-Toth, I.I.K., \& Nauber, U., 1979}{84}{942}
\ref\apj{Wrobel, J.M., \& Krause, S.W., 1990}{363}{11}
\ref\apjs{Xu, W., Readhead, A.C.S., Pearson, T.J., Polatidis, A.G., \& Wilkinson, P.N., 1995}
  {99}{297}

\end{references}
\end{document}